\documentclass[preprint2]{aastex}

\newcommand{\muK}{\mu{\rm K}}
\newcommand{\C}{{\cal C}}
\newcommand{\calN}{{\cal N}}
\newcommand{\calD}{{\cal D}}
\newcommand{\etal}{{et al.}}

\slugcomment{CfPA-98-th-16; CITA-98-25; astro-ph/9808264}
\shortauthors{Bond, Jaffe, \& Knox}
\shorttitle{CMB Data Compression}

\begin{document}

\title{Radical Compression of\\
Cosmic Microwave Background Data}
\author{J.~R.~Bond\altaffilmark{1}}
\affil{Canadian Institute for Theoretical Astrophysics, Toronto, ON M5S
  3H8, CANADA}
\author{A.~H.~Jaffe\altaffilmark{2}}
\affil{Center for Particle Astrophysics,
  301 LeConte Hall, University of California, Berkeley, CA 94720}
\and
\author{L.~Knox\altaffilmark{3}}
\affil{Canadian Institute for Theoretical Astrophysics, Toronto, ON M5S
  3H8, CANADA}
\altaffiltext{1}{{\tt bond@cita.utoronto.ca}}
\altaffiltext{2}{{\tt jaffe@cfpa.berkeley.edu}}
\altaffiltext{3}{{\tt knox@flight.uchicago.edu}; Current address:
  Department of Astronomy and Astrophysics, 5640 S.\ Ellis Ave.,
  Chicago, IL 60637} 
\begin{abstract}
  Powerful constraints on theories can already be inferred from {\it
    existing}\/ CMB anisotropy data.  But performing an exact analysis
  of available data is a complicated task and may become prohibitively
  so for upcoming experiments with $\gtrsim10^4$ pixels.  We present a
  method for approximating the likelihood that takes power spectrum
  constraints, {\sl e.g.}, ``band-powers'', as inputs.  We identify a bias
  which results if one approximates the probability distribution of the
  band-power errors as Gaussian---as is the usual practice.  This bias
  can be eliminated by using specific approximations to the
  non-Gaussian form for the distribution specified by three parameters
  (the maximum likelihood or mode, curvature or variance, and a third
  quantity). We advocate the calculation of this third quantity by
  experimenters, to be presented along with the maximum-likelihood
  band-power and variance.  We use this non-Gaussian form to estimate
  the power spectrum of the CMB in eleven bands from multipole moment
  $\ell = 2$ (the quadrupole) to $\ell=3000$ from all published
  band-power data.  We investigate the robustness of our power spectrum
  estimate to changes in these approximations as well as to selective
  editing of the data.
\end{abstract}


\section{Introduction}

Measurement of the anisotropy of the Cosmic Microwave Background (CMB)
is proving to be a powerful cosmological probe. Proper statistical
treatment of the data---likelihood calculation---is complicated and
time-consuming, and promises to become prohibitively so in the very near
future.  Here, we introduce approximations for this likelihood
calculation which allow simple and accurate evaluation after the
direct estimation of the power spectrum ($C_\ell$) from the data.

Although it is possible to produce constraints on cosmological
parameters directly from the data, using the power spectrum as an
intermediate step ({\sl e.g.} \citet{tegmark}) has several advantages.  
The near-degeneracy of some
combinations of cosmological parameters ({\sl e.g.}, \citet{BET97}) implies the
surfaces of constant likelihood in cosmological parameter space are
highly elongated, making it difficult for search algorithms to
navigate \citep{OhSpergelHinshaw}.  The power spectrum space is much
simpler than the cosmological parameter space since each multipole
moment (or band of multipole moments) is usually only weakly dependent
on the others, alleviating the search difficulties.  Although one
still has the problem left of estimating nearly degenerate
cosmological parameters from the resulting power spectrum constraints,
the likelihood given the power spectrum constraints is much easier to
compute than the likelihood given the map data.

Proceeding via the power spectrum also facilitates the calculation of
constraints from multiple datasets.  Without this intermediate step, a
joint analysis may often be prohibitively complicated.  Aspects
particular to each experiment ({\sl e.g.}, offset removals, non-trivial
chopping strategies) make implementation of the analysis sufficiently
laborious that no one has jointly analyzed more than a handful of
datasets in this manner.  Reducing each dataset to a set of constraints
on the power spectrum can serve as a form of data compression which
simplifies further analysis.  Indeed, most studies of cosmological
parameter constraints from all, or nearly all, of the recent data have
used, as their starting points, published power spectrum constraints,
[\citep{lineweaver97,lineweaver98a,lineweaver98b,lineweaver98c,lineweaverBarbosa97,hancockrocha};
see also \citet{tegmark98} and \citet{efstetal98} for joint analyses
with other datasets].  Since the power spectrum constraints are usually
described with orders of magnitude fewer numbers than the pixelized
data, we refer to this compression as ``radical''.

Are there any disadvantages to proceeding via the power spectrum?  To
answer this question, let us consider the analysis procedure.  Most
analyses of CMB datasets have assumed the noise and signal to be
Gaussian random variables, and to date there is no strong evidence to
the contrary (although for a different view, see \citet{weird16}).
The simplicity of this model of the data allows for an
exact Bayesian analysis, which has been performed for almost all
datasets individually.  The procedure is conceptually straightforward:
maximize the probability $P({\rm parameters} | {\rm data})$ over the
allowed parameter space.  Most often, we take the prior probability
for the parameters to be constant, so this is equivalent to maximizing
the likelihood, $P({\rm data}|{\rm parameters})$. Because we have
assumed the noise and signal to be Gaussian, this latter is just a
multivariate Gaussian {\em in the data}; the theoretical parameters
enter into the covariance matrix.

Fortunately, if the theoretical signal is indeed normally distributed
and in addition the signals are statistically isotropic, the power
spectrum encodes all of the information about the model, and all of
the constraints on the parameters of the theory can be obtained from
the $C_\ell$ probability distribution: {\sl i.e.}, the likelihood as a
function of some (cosmological) parameters, $a_i$, is just the
likelihood as a function of the power spectrum determined from those
parameters: $P({\rm data}|a_i)=P({\rm data}|C_\ell[a_i])$.  Thus the
constraints on the power spectrum may serve as our ``compressed
dataset''. (If the theory is not isotropic, as may occur for
nontrivial topologies ({\sl e.g.}, \citet{BPS98}), or is non-Gaussian, then the
analysis must go beyond the isotropic power spectrum.)

A problem arises though due to the fact that the uncertainties in the
power spectrum determination are not Gaussian-distributed.  Thus if we
compress the power spectrum probability distribution to a mean (or
mode---the location of the posterior maximum) and a variance, we lose
the information contained in the higher order moments.  One might be
tempted to rely on the central limit theorem and hope that the posterior
for the power spectrum is sufficiently close to a Gaussian that a simple
$\chi^2$ procedure will suffice.  This is what has been done in recent
joint analyses of current CMB data ({\sl e.g.},
\citet{lineweaver97}; \citet{lineweaver98a}; \citet{lineweaver98b};
\citet{lineweaver98c}; \citet{lineweaverBarbosa97}; \citet{hancockrocha}).
and what has been advocated for the analysis of satellite data ({\sl
  e.g.}, \citet{tegmark}).

Not only is information discarded with this procedure, however---which
one might think of as merely increasing the final error bars---but
neglect of this effect leads to a bias
\citep{jkbtexas,bjkpspec,uroscopy,OhSpergelHinshaw}.  Here we show that
the information loss and its effects, such as this bias, can be greatly
reduced by assuming the posterior distribution to have a specific
non-Gaussian form parameterized by the likelihood maximum, the
covariance matrix and a third quantity which measures the noise
contribution to the uncertainty of each measured amplitude.

A relatively fast algorithm for determining the power spectrum or other
parameters is the ``quadratic estimator''
\citep{tegmark,bjkpspec,OhSpergelHinshaw}, although it still requires
$O(n^3_{\rm pix})$ operations, where $n_{\rm pix}$ is the number of
pixels in the dataset. Our view of quadratic estimation is that, used
iteratively, it is a particular method for finding the maximum of the
likelihood and the parameter covariance matrix
\citep{bjkpspec,OhSpergelHinshaw}.  We emphasize that the information
loss associated with compression to a mode and covariance matrix has
nothing to do with how that mode and covariance matrix are calculated.
Other methods of likelihood analysis will, of course, suffer the same
problems when the constraints are reduced to these two sets of
quantities.  In fact, the quadratic estimation algorithm has the
advantage that an implementation of it as a computer code can be used
(with very minor changes) to calculate the new noise contribution
quantity.

In Section~\ref{sec:likelihood}, we describe the problems generated by
the non-Gaussianity of the likelihood function, and propose solutions
which allow rapid and simple calculation of cosmological likelihoods for
CMB data at the price of calculating only this single new parameter at
each $\ell$ (or band). \citet{bartlettetal} have used an ansatz,
essentially equivalent to the equal-variance approximation of
Sec.~\ref{sec:indmode} to arrive at similar results.) In
Section~\ref{sec:dmr}, we test this method via application to COBE/DMR
data. In Section~\ref{sec:general}, we extend the formalism to more
complicated chopping experiments and to the measurement of other
amplitude parameters such as bandpowers. We apply
these extensions to the Saskatoon data in Section~\ref{sec:SK}, to OVRO,
SP and SuZIE in Section~\ref{sec:upperlim}, and Saskatoon combined with
COBE/DMR, in Section~\ref{sec:results}.

We then apply our procedure to the more ambitious task of fitting an
eleven parameter model to a compendium of all CMB results to date in
Section~\ref{sec:bands}.  Previous explorations of parameter space have
been limited to much lower dimensionality and have assumed Gaussianity
({\sl e.g.},
\citet{lineweaver97}; \citet{lineweaver98a}; \citet{lineweaver98b};
\citet{lineweaver98c}; \citet{lineweaverBarbosa97}; \citet{hancockrocha}).
The parameters are the power in eleven bins from $\ell = 2$ to
$\ell=3000$.  We study the robustness of the resulting maximum
likelihood power spectrum to assumptions about the noise contribution to
the error, different binnings and selective editings of the data.  The
binned power spectrum, fit to the band-power data, provides an excellent
tool for visualizing the combined power spectrum constraints from all
the data.  Such a figure should replace the usual one of all the band
powers, which is much harder to interpret.

Finally, in Section~\ref{sec:conclude} we
discuss the results and conclude, ending with an exhortation to the
community to calculate and provide the appropriate quantities for all
future experiments.

Throughout, we will use $C_\ell$ to refer to the usual CMB power
spectrum, and define
\begin{equation}
  {\cal C}_\ell\equiv {\ell(\ell+1)\over2\pi}C_\ell.
\end{equation}

\section{Non-Gaussianity of the Likelihood Function}
\label{sec:likelihood}
\subsection{The Problem: Cosmic Bias}
\label{sec:bias}

To illustrate the problem, consider an unrealistic ``experiment''
covering the whole sky with no noise.  In this case, the data at pixel
$p$, $\Delta_{p}$ is just the actual sky signal, $s_{p}$, and the
correlation matrix, $S_{pp'}$, is just the correlation function
$c(\theta_{pp'})$.  Even in this case, the observed sky is just one
realization of the underlying power spectrum.  To determine these
$\C_\ell\equiv \ell(\ell+1)C_\ell/(2\pi)$, we still must resort to the
likelihood function.  In this case,
\begin{eqnarray}
  \label{eqn:nonoiselike}
  -2\ln P(\Delta|\C_{\ell})&=&\ln\det S(\C_{\ell}) + \Delta^\dagger
  S^{-1} \Delta\nonumber\\
  &=& \sum_{\ell}(2\ell+1)\left(\ln \C_{\ell} + {\widehat
    \C}_{\ell}/\C_{\ell}\right)\, ,
\end{eqnarray}
up to an irrelevant additive constant.  In the second line, we define
the observed power spectrum of this realization as
\begin{equation}
{\widehat \C}_{\ell}={\ell(\ell+1)\over2\pi}\
{1\over2\ell+1}\sum_{m}|a_{\ell m}|^{2}\, ,
\end{equation}
where the $a_{\ell m}$ are the spherical harmonic coefficients of the
(noise-free) observed sky.

A Gaussian distribution has the following properties: it is completely
specified by its mean and covariance matrix (higher moments of the
distribution can be derived from these); the covariance matrix is given
by the inverse of the curvature matrix (defined as ${\cal
  F}_{\ell\ell'}=-{\partial^2 \ln P(\Delta|\C_{\ell})/\partial
  \C_\ell\partial \C_{\ell'}}$); and the curvature matrix is independent
of $\C_\ell$.  None of these properties hold for the distribution in
Eq.~\ref{eqn:nonoiselike}, which is non-Gaussian.\footnote[1]{The
  posterior distribution of $\C_\ell$ is not $\chi^2_{2\ell+1}$ either.
  It is the realization, ${\widehat \C}_{\ell}$, that is
  $\chi^2_{2\ell+1}$-distributed for a fixed ``underlying'' power
  spectrum, $\C_\ell$.}  
For a non-Gaussian distribution, it is
therefore often useful to consider the ensemble average curvature, also
known as the {\em Fisher Matrix}, $F\equiv\langle{\cal F}\rangle$.

What happens if, despite the non-Gaussianity of the
distribution in Eq.~\ref{eqn:nonoiselike}, one identifies
the covariance matrix with the inverse of the curvature matrix
and then ignores higher order moments?  The first step would
be calculation of the curvature matrix:
\begin{equation}
  -{\partial^2 \ln P(\Delta|\C_{\ell})\over\partial \C_\ell\partial
    \C_{\ell'}} =
  {2\ell+1\over2}\left(2{\widehat \C}_\ell/\C_\ell^3-1/\C_\ell^2\right)\delta_{\ell\ell'}.
\end{equation}
Note that, unlike for a Gaussian distribution, the curvature depends on
$\C_\ell$.  The natural remedy is to evaluate it at the peak of the
likelihood, ${\widehat \C}_\ell$.  The standard error (square root of
the variance) is then given by $\delta \C_\ell =
\sqrt{2/(2l+1)}\C_\ell$.  Note though that, uncertainties derived in
this manner are larger if ${\widehat \C}_\ell$ has fluctuated upward
from the underlying ``real'' value and smaller for a downward
fluctuation.  If, in addition, we ignore higher order moments of the
distribution, then upward fluctuations are given less weight than
downward fluctuations, resulting in a downward bias for the overall
power spectrum amplitude.  It is generally the lowest multipole moments
constrained by an observation that have the most non-Gaussian
distributions.  As has been seen \citep{BunnWhite,bjkpspec} and 
will be seen again below, this may contribute to some of the confusion in
the community regarding the so-called anomalous value of the COBE
quadrupole.

\begin{figure}
  \plotone{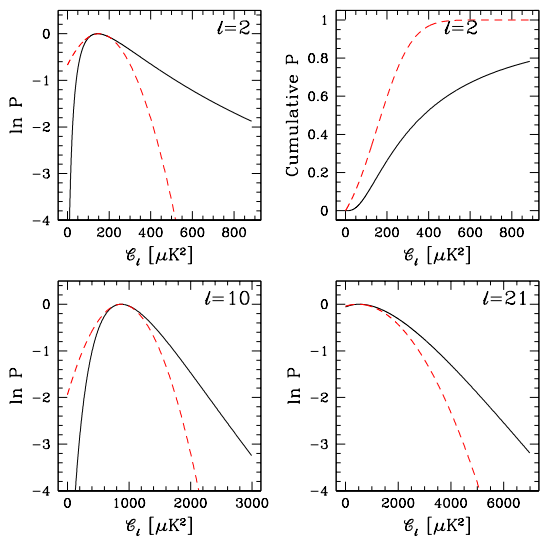}
  \caption{DMR Likelihoods $P(\Delta|\C_\ell$) for
    various values of $\ell$, as marked. The horizontal axis is ${\cal
      C}_\ell=\ell(\ell+1)C_\ell/(2\pi)$. The upper right panel gives
    the cumulative probability. The solid (black) line is the full
    likelihood calculated exactly. The dashed (red) line is the Gaussian
    approximation about the peak.\label{fig:Cl_like1}}
\end{figure}

In the presence of noisy data over partial areas of the sky, the
likelihood is no longer so simple, and must be laboriously calculated
({\sl e.g.}, \citet{Bond94};\citet{BunnWhite};\citet{bondjaffe1};\citet{bondjaffe2}). In
Fig.~\ref{fig:Cl_like1} we show the actual likelihood for the COBE
quadrupole and other multipoles, along with the Gaussian that would be
assumed given the curvature matrix calculated from the data.  The
figures show another way of understanding the bias introduced by
assuming Gaussianity: upward deviations from the mean (which is not
actually the mean of the non-Gaussian distribution, but the mode) are
overly disfavored by the Gaussian distributions while downward ones are
overly probable.  For example, the standard-CDM value of
${\C}_2=770\muK^2$ is only 0.2 times less likely than the most likely
value of $150\muK^2$ but it seems like a 5-sigma excursion ($4\times
10^{-6}$ times less likely) based on the curvature alone.

Although it is extremely pronounced in the case of the quadrupole this
is a problem that plagues all CMB data: the actual distribution is
skewed to allow larger positive excursions than negative.  The full
likelihood ``knows'' about this and in fact takes it into account;
however, if we compress the data to observed $\C_\ell\pm\sigma_\ell$ (or
even observed $\C_\ell$ and a correlation matrix $M_{\ell\ell'}$) we
lose this information about the shape of the likelihood function.
Because of its relation to the well-known phenomenon of cosmic variance,
we choose to call this problem one of {\em cosmic bias}.

We emphasize that cosmic bias can be important even in high-S/N
experiments with many pixels. We might expect the central limit theorem
to hold in this case and the distributions to become Gaussian. Indeed
they do, at least near the peak. However, the central limit theorem does
not guarantee that the tails of the distribution will approach those of
a Gaussian as rapidly as does the region near the peak and there is the
danger that a few seemingly discrepant points are given considerably
more weight than they deserve. Cosmic bias has also been noted in
previous work \citep{bjkpspec,uroscopy,OhSpergelHinshaw}.

Putting the problem a bit more formally, we see that even in the limit
of infinite signal-to-noise we cannot use a
simple $\chi^2$ test on $\C_\ell$ estimates; such a test implicitly
assumes a Gaussian likelihood. Unlike the distribution discussed here, a
Gaussian would have constant curvature ($\delta \C_\ell={\rm constant}$),
rather than $\delta \C_\ell\propto \C_\ell$ as illustrated here.

We emphasize that the problem as outlined here is easily solved in
principle: just calculate using the full likelihood function.
Unfortunately, this is more easily said than done---direct calculation
of the likelihood function takes $O(n_p^3)$ operations per
parameter-space point. For the $n_p\gtrsim10^5$ datasets already coming,
this is prohibitively expensive. Indeed, it is not even clear how to
perform the operations necessary even to find the likelihood peak and
variance in a reasonable time \citep{bjkpspec,OhSpergelHinshaw}. It is
likely that other forms of data compression and/or new algorithms will
be necessary even at this stage of the analysis. Signal-to-noise
Eigenmodes, discussed in Appendix~\ref{app:snmode}, have been suggested
as a useful compression tool \citep{TTHklmode,jkbtexas} and used in
some analyses \citep{bondjaffe1,bondjaffe2,BunnSugiyama,BunnWhite}.

Instead, we must find efficient ways to approximate the likelihood
function based on minimal information. In the rest of this paper, we
discuss two approximations, each motivated by different aspects of our
knowledge of the likelihood function. Each requires only knowledge of
the likelihood peak and curvature (or variance) as well as a third
quantity related to the noise properties of the experiment. Alternately,
for already-calculated likelihood functions, each approximation gives a
functional form for fitting with a small number of parameters.

\subsection{The solution: approximating the likelihood}
\label{sec:solution}
\subsubsection{Offset lognormal distribution}
We already know enough about the likelihood to see a solution to this
problem. For a given multipole $\ell$, there are two distinct regimes of
likelihood. Add uniform pixel noise and a finite beam to the simple all-sky
``experiment'' considered above. Now, the likelihood has contributions
from the signal, $a_{\ell m}$, and the noise, $n_{\ell m}$ (after
transforming again to spherical harmonics).
\begin{eqnarray}
  \label{eqn:fullskylike}
  &&-2\ln P(\Delta|\C_{\ell})=\nonumber\\
&&\sum_{\ell}
  (2\ell+1) \left[
    \ln\left(\C_\ell B_\ell^2 + \calN_\ell\right) 
    +{{\widehat \calD}_{\ell}\over\C_\ell B_\ell^2 + \calN_\ell}\right]
\end{eqnarray}
(as usual up to an irrelevant additive constant),
with $\calN_\ell=\ell(\ell+1)N_\ell/(2\pi)$, where $N_\ell=\langle|n_{\ell m}|^2\rangle$ is the noise power spectrum in spherical harmonics, and
${\widehat \calD}_{\ell} \equiv [\ell(\ell+1)/(2\pi)]\ \sum_m |a_{\ell
  m}|^2/(2\ell+1)$ is the power spectrum of the full data (noise plus
beam-smoothed signal); we have written it as a different symbol from
above to emphasize the inclusion of noise and again use script lettering
to refer to quantities multiplied by $\ell(\ell+1)/(2\pi)$.

Now, the likelihood is maximized at
$\C_\ell=({\widehat\calD}_\ell-\calN_\ell)/B_\ell^2$ and the curvature
about this maximum is given by
\begin{equation}\label{eqn:fullcurv}
  {\cal F}^{(\C)}_{\ell\ell'}=
  -{\partial^2 \ln P(\Delta|\C_{\ell})\over\partial \C_\ell\partial
    \C_{\ell'}} =
  {2\ell+1\over2}\left(\C_\ell+\calN_\ell/B_\ell^2 \right)^{-2}\delta_{\ell\ell'}
\end{equation}
so the error (defined by the variance) on a $\C_\ell$ is
\begin{equation}
  \delta \C_\ell = \left(\C_\ell+\calN_\ell/B_\ell^2 \right)/\sqrt{\ell+1/2}.
\end{equation}

Note that in this expression there is once again indication of a bias
if we assume Gaussianity: upward fluctuations have larger uncertainty
than downward fluctuations.  But this is not true for $Z_\ell$ where
$Z_\ell$ is defined so that $\delta Z_\ell \propto \delta \C_\ell /
(\C_\ell + \calN_\ell/B_\ell^2)$.  More precisely, $Z_\ell \equiv
\ln(\C_\ell + \calN_\ell/B_\ell^2)$.  Since $\delta Z_\ell$ is
proportional to a constant, our approximation to the likelihood is to
take $Z_\ell$ as normally distributed.  That is, we approximate
\begin{equation}\label{eqn:gausslike}
 -2\ln P(\Delta|\C_{\ell})=\sum_{\ell\ell'} Z_\ell
M^{(Z)}_{\ell\ell'} Z_{\ell'}
\end{equation}
(up to a constant) where $M^{(Z)}_{\ell\ell'} = (\C_\ell+x_\ell)M^{(\C)}_{\ell\ell'}
(\C_{\ell'}+x_{\ell'})$ where $M^{(\C)}$ is the weight matrix (covariance
matrix inverse) of the $\C_\ell$, usually taken to be 
the curvature matrix.
We refer to Eq.~\ref{eqn:gausslike} as the offset lognormal distribution
of $\C_\ell$. 
Somewhat more generally we write
\begin{equation}\label{eqn:Zl}
  Z_\ell = \ln(\C_\ell + x_\ell)
\end{equation}
for some constant $x_\ell$, which for the case at hand is
$x_\ell=\calN_\ell/B_\ell^2$.

It is illustrative to derive the quantity $Z_\ell$ in a somewhat more
abstract fashion. We wish to find a change of variables from $\C_\ell$
to $Z_\ell$ such that the curvature matrix is a constant:
\begin{equation}
  {\partial {\cal F}^{(Z)}_{\ell\ell'}\over\partial Z_\ell}=0.
\end{equation}
That is, we want to find a change of variables such that
\begin{equation}\label{eqn:changevar}
  \left({\cal F}^{(Z)}\right)^{-1}_{LL'} = \sum_{\ell\ell'}
{\partial Z_L\over\partial \C_\ell} \left({\cal F}^{(\C)}\right)^{-1}_{\ell\ell'}
{\partial Z_{L'}\over\partial \C_{\ell'}}
\end{equation}
is not a function of $Z$. We immediately know of one such transformation
which would seem to do the trick:
\begin{equation}
\label{eqn:dotrick}
  {\partial Z_L\over\partial \C_\ell} = {\cal F}_{L\ell}^{1/2}
\end{equation}
where the ${1/2}$ indicates a Cholesky decomposition or Hermitian square
root. In general, this will be a horrendously overdetermined set of
equations, $N^2$ equations in $N$ unknowns. However, we can solve this
equation in general if we take the curvature matrix to be
given everywhere by the diagonal
form for the simplified experiment we have been
discussing (Eq.~\ref{eqn:fullcurv}).  In this case, the equations decouple and lose their
dependence on the data, becoming (up to a constant factor)
\begin{equation}
  {dZ_\ell\over d\C_\ell} = \left(\C_\ell+\calN_\ell/B_\ell^2\right)^{-1}.
\end{equation}
The solution to this differential equation is just what we expected,
\begin{equation}
  Z_\ell=\ln(\C_\ell+\calN_\ell/B_\ell^2)
\end{equation}
with correlation matrix
\begin{equation}
  \left({\cal F}^{(Z)}\right)^{-1}_{\ell\ell'}=
  {\left({\cal F}^{(\C)}\right)^{-1}_{LL'}\over
  \left(\C_L+\calN_L/B_L^2\right)\left(\C_{L'}+\calN_{L'}/B_{L'}^2\right)}
\end{equation}
where $\ell = L$ and $\ell'=L'$.
Please note that we are calculating a constant correlation matrix; the
$\C_\ell$ in the denominator of this expression should be taken at the
peak of the likelihood ({\sl i.e.}, the estimated quantities).

We emphasize that, even for an all-sky continuously and uniformly
sampled experiment (for which Eq.~\ref{eqn:fullcurv} is exact), this
Gaussian form, Eq.~\ref{eqn:gausslike}, is only an approximation, since
the curvature matrix is given by Eq.~\ref{eqn:fullcurv} only at the peak.
Nonetheless we expect it to be a better approximation than a naive
Gaussian in $\C_\ell$ (which we note is the limit $x_\ell\to\infty$ of
the offset lognormal).

We also note that there is another approximation involved in this
expression: by choosing just a single independent change of variable for
each $\ell$ we assume that the correlation structure is unchanged as you
move away from the likelihood peak.

Often a very good approximation to the curvature matrix is its ensemble
average, the Fisher matrix, $F \equiv \langle {\cal F} \rangle$.  Below,
unless mentioned otherwise, we use the Fisher matrix in place of the
curvature matrix.  However, we will see that in our application to the
Saskatoon data, the differences between the curvature matrix and Fisher
matrix can be significant.

\subsubsection{The Equal Variance Approximation} \label{sec:indmode}

In this subsection we consider an alternate form for the likelihood
function ${\cal L}\propto P(\Delta|\C_{\ell})$ that is sometimes
a better approximation than the offset-lognormal form.  The
approximation is exact in the limit that the observations
can be decomposed into modes that are independent, with equal
variances.  For example, for a switching experiment in which
the temperature of $G$ pixels are measured, with the same noise
$\sigma_N$ at each point, and such that each pixel is far enough from
the others that there is no correlation between the points, 
the likelihood can be written as
\begin{equation}
\label{eqn:equalindepmode}
\ln{\cal L}=-G/2\left[ \ln \left(\sigma_T^2 + \sigma_N^2\right) +
{\sum_i d_i^2/G \over \sigma_T^2 + \sigma_N^2 }\right]
\end{equation}
where $\sigma_T^2$ is such that $C_{T,ij} = \sigma_T^2 \delta_{ij}$
and $d_i$ are the pixel temperatures.  The independent pixel idealization
was very close to the case for the OVRO experiment, and, as we
show in Section~\ref{sec:upperlim}, the calculated likelihood
is well approximated by this equation.  The maximum likelihood occurs at
a signal amplitude ${\widehat\sigma}_{\rm T}$
which is related to the data by 
${\widehat\sigma}_{\rm T}^2=\sum d_i^2/G-\sigma_N^2$.

\begin{figure}   
  \plotone{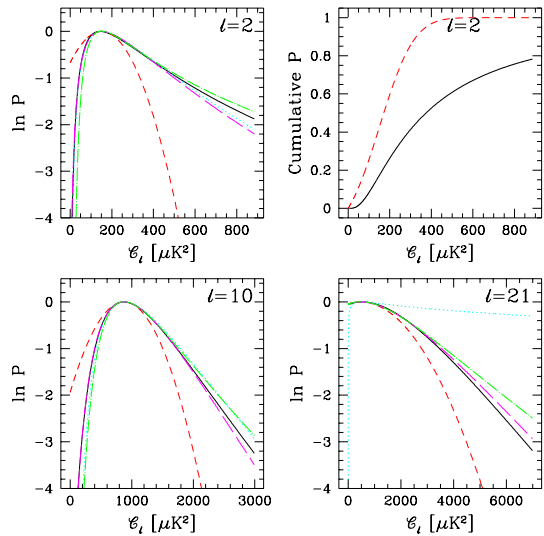}
  \caption{Full and approximate COBE/DMR likelihoods
    $P(\Delta|\C_\ell)$ for various values of $\ell$, as marked. The
    horizontal axis is ${\cal C}_\ell=\ell(\ell+1)C_\ell/(2\pi)$. The
    upper right panel gives the cumulative probability. The solid (black)
    line is the full likelihood calculated exactly. The short-dashed (red)
    line is the Gaussian approximation about the peak. The dotted (cyan)
    line is a Gaussian in $\ln{\C_\ell}$; the dashed (magenta) line is a
    Gaussian in $\ln{(\C_\ell+x_\ell)}$, as discussed in the text. The
    dot-dashed (green) line is the equal-variance approximation.
    \label{fig:dmrlike}}
\end{figure}

If we define $Z = \ln\left(\sigma_T^2 + x\right)$, 
$\widehat Z  = \ln\left(\widehat \sigma_T^2 + x\right)$, and
$x=\sigma_N^2$ then we can rewrite
Eq.~\ref{eqn:equalindepmode} in a form that will be useful for relating it to
the
previous offset-
lognormal form:
\begin{eqnarray}
  \label{eqn:equalindepmode2}
  \ln {\cal L}/{\widehat{\cal L}}&=& -{G\over 2}
  \left[e^{-(Z-{\widehat Z})}-\left(1-(Z-{\widehat Z})\right)\right].
\end{eqnarray}
Note that if we consider only a single $\ell$ then the above form applies
to Eq.~\ref{eqn:fullskylike} as well, with $G=2\ell+1$ and $Z=\ln
\left(\C_\ell+x_\ell\right)$.  We know this should be the case since the
likelihood of Eq.~\ref{eqn:fullskylike} (for a single $\ell$) is also
one for independent modes ($a_{lm}$) with equal variances ($\C_\ell +
x_\ell$). If we fix $G$ and $x$ for each mode ({\sl e.g.}, band of $\ell$), we
refer to this as the ``equal variance approximation.''  Also note that
the first term in the expansion of Eq.~\ref{eqn:equalindepmode2} in
$Z-\widehat Z$ is $-G/2\left(Z-\widehat Z\right)^2$, which with the
identification, $G=2{\cal F}^{(Z)}$, is the offset-lognormal form.  Thus
when the modes have equal variance and are independent, then the offset
lognormal form is simply the first term in a Taylor expansion of the
equal-variance form.  An advantage of the full form is that the
asymptotic form $-(G/2)(Z-{\widehat Z})$ linear in $Z$ for large signal
amplitudes holds (and thus gives a power law rather than exponential
decay in ${\cal L}$), whereas the offset lognormal is dominated by the
$Z^2$.  An advantage of the offset-lognormal form is that it does not
require the existence of equal and independent modes.
Figure~\ref{fig:dmrlike} shows that for the range of relevance for the
likelihoods for DMR, the offset lognormal and the equal/independent
variance likelihood approximations are quite close over the dominant
2-sigma falloff from maximum. We have found this to generally be true.

For either form, three quantities need to be specified, the
noise-related offset $x$, ${\widehat Z}$ and $G$. Given $x$, ${\widehat
  Z}$ is determined from the maximum likelihood and $G$ can be
determined from the curvature of the likelihood. One could also specify
the amplitudes $\C$ at three points, {\sl e.g.}, at the maximum and the places
where ${\cal L}/{\widehat{\cal L}}$ falls by $e^{-1/2}$, the upper and
lower one-sigma errors if the distribution were fit on either side by a
Gaussian. Forcing the approximation to pass through these points
enforces values of $x$, ${\widehat Z}$ and $G$.

In Section~\ref{sec:bands}, we apply these approximations to power
spectrum
estimation from current data for which the practice has been to quote a
signal amplitude with upper and lower one-sigma errors, say
${\widehat\C}$, $\C_u$ and $\C_d$. Often these are Bayesian estimates,
determined by choosing a prior probability for $\C$ and integrating the
likelihood. Sometimes the $e^{-1/2}$ points are given, which are
slightly easier to implement in fitting for $x$ and $G$. Since the tail
of Eq.~\ref{eqn:equalindepmode2} is quite pronounced, resulting in a
dramatic asymmetry in ${\cal L}$ between the up and down sides of the
maximum even in the $Z$ variable, we have found that just using the
second derivative of the likelihood or the Fisher matrix approximation
to it to fix $G$ is not as good as assuming the offset lognormal and requiring
that the functional forms match at the upper $e^{-1/2}$ point. Thus, if
the error $\sigma_\C= 1/\sqrt{{\cal F}^{(\C)}}$ is from the curvature or
Fisher matrix, then we prefer the choice
\begin{equation}
\label{eqn:chi2approxparams}
G=\left[e^{-\sigma_Z}
-(1-\sigma_Z)\right]^{-1} \, , \  \sigma_Z
={\sigma_\C \over {\widehat\C}+x} = {1\over \sqrt{{\cal F}^{(Z)}}}
\end{equation}
rather than the curvature form $G=2/\sigma_Z^2$, or the $[{\rm
cosh}(\sigma_Z) -1]^{-1}$ average of the $\pm 1/2$ widths.  This is
what was done in Fig.~\ref{fig:dmrlike}, and in all subsequent
figures.

Fig.~\ref{fig:dmrlike} shows that a linear $-GZ/2$ asymptote in the
log-likelihood is not always correct and sometimes the lognormal does
better. That the tail often declines faster can be understood in terms
of an effective number of modes $G(\C^{-1})$ which increases as $\C$
increases. To demonstrate this, it is useful to consider the likelihood
behavior for ``signal-to-noise'' eigenmodes, which are linear
combinations of the pixelized data which make them statistically
independent for Gaussian signals and noise.  The data is then
characterized by observed amplitudes $d_k$, with a noise contribution
transformed to give unity variance, and a signal contribution with
amplitude $\C \lambda_k$, in terms of a ``signal-to-noise'' eigenvalue
$\lambda_k$ and an overall amplitude $\C$. Such transformations have
been much discussed in the literature ({\sl e.g.},
\citet{Bond94};\citet{BunnSugiyama};\citet{BunnWhite};\citet{bjkpspec}), and we will not go into the details
here; for more details on using this formalism to examine the overall
form of the likelihood function, see Appendix~\ref{app:snmode}.

The cases in which we expect Eq.~\ref{eqn:equalindepmode2} to be a
good approximation are those in which the eigenmodes have a broad region
over which $\lambda$ varies slowly and a very rapid falloff towards zero
beyond. This is exact for the independent pixel points described above,
with $ \lambda = \sigma_T^2/\sigma_{N}^2$ the same for all $G$
modes. If there were a number of frequency channels as well as pixels,
only the linear combinations which are flat in thermodynamic temperature
have this $\lambda$, and the rest are zero.  For cases where the
equal variance approximation is not exact, the signal-to-noise modes
will have different eigenvalues $\lambda$.  One might then take an effective
$G$ to be the number of modes with $\lambda>1$ or some other cutoff 
(since these
are the ones that have greater signal than noise).  However, then $G$
grows as $\C_\ell$ increases, altering the power-law tail.

Thus, the very general approach of ``signal-to-noise'' eigenmodes has
allowed us to understand that the simple law with an effective $G$
will usually overshoot the high $\C$ tail somewhat, even though it
fits very well to 1-sigma, and usually beyond. The offset lognormal form,
motivated by it, could err on either side, since it would presuppose a
specific sort of increase in the number of eigenmodes
contributing. Fortunately either approximation seems to work well
enough to allow accurate parameter estimation from a very small set of
numbers.

\section{Application to COBE/DMR}
\label{sec:dmr}

We first apply these methods to the anisotropy measurements of the DMR
instrument on the COBE satellite \citep{DMR}. The DMR instrument
actually measured a complicated set of temperature differences
$60^\circ$ apart on the sky, but the data were reported in the much
simpler form of a temperature map, along with appropriate errors (which
we have expanded to take into account correlations generated by the
differencing strategy, as treated in \citet{Bond94}, following
\citet{LineSmoot}.  The calculation of the theoretical correlation
matrix includes the effects of the beam, digitization of the time
stream, and an isotropized treatment of pixelization, using the table
given by \citet{GneSmoot}, modified for resolution 5.  We use a weighted
combination of the 31, 53 and 90~GHz maps. Because most of the
information in the data is at large angular scales, we use the maps
degraded to ``resolution 5'' which has 1536 pixels. Further, we cannot
of course observe the entire CMB sky; we use the most recent galactic
cut suggested by the COBE/DMR team \citep{DMR}, leaving us with 999
pixels to analyze.  We use the galactic, as opposed to ecliptic,
pixelization.

Before we can apply our procedure to COBE/DMR, we must discuss how to deal
with the partial sky coverage of any real CMB experiment.
To a good approximation, the COBE/DMR Fisher
matrix can be written as ({\sl e.g.},
\citet{Knox95};\citet{Jungmanetal};\citet{bjkpspec};\citet{HobsonMagueijo}) 
\begin{equation}
  \label{eqn:partialsky_fish}
  F^{(\C)}_{\ell\ell'}=f_{\rm sky}{2\ell+1\over2}\left[
    \C_\ell + {\ell(\ell+1)\over2\pi wB_\ell^2}\right]^{-2}\delta_{\ell\ell'}.
\end{equation}
In the language of our new procedure, this means that we still expect
to be able to approximate the likelihood as a Gaussian in the same
$Z_\ell=\ln(\C_\ell+x_\ell)$, but now we can only approximate the term
$x_l\simeq \ell(\ell+1)/[2\pi w B^2(\ell)]$, where $w$ is the weight
per solid angle of the experiment. In terms of the total weight, $W$,
of the experiment, $w=W/(4\pi f_{\rm sky})$. A more detailed
approximation for a particular experiment might be possible, but as we
will see below, this expression does extremely well in reproducing the
full non-Gaussian likelihood.

We have calculated the maximum-likelihood power spectrum and its error
(Fisher) matrix using the quadratic estimator procedure of
\citep{bjkpspec}. With knowledge of the COBE/DMR beam \citep{DMR}
along with the noise properties of the experiment, we can calculate the
necessary quantity $x_\ell$. For COBE/DMR, we have an average inverse
weight per solid angle of $w^{-1} = 9.5 \times 10^{-13}$ (equivalent to
an RMS noise of $22\;\mu{\rm K}$ on $7^\circ\times7^\circ$ pixels).

With these numbers, we show the full likelihood in comparison to the
``naive Gaussian'' approximation, as well as our offset lognormal
ansatz.  While the naive Gaussian approximation consistently
overestimates the likelihood below the peak and underestimates it above
the peak, the lognormal form reproduces the full expression extremely
well in both regimes.

The Gaussian form of the offset lognormal form makes using the power
spectrum estimates for parameter estimation very simple: we evaluate a
$\chi^2$ in the quantity $Z_\ell$ rather than $\C_\ell$ (although the
model is now nonlinear in the spectral parameters).

\begin{figure}
  \plotone{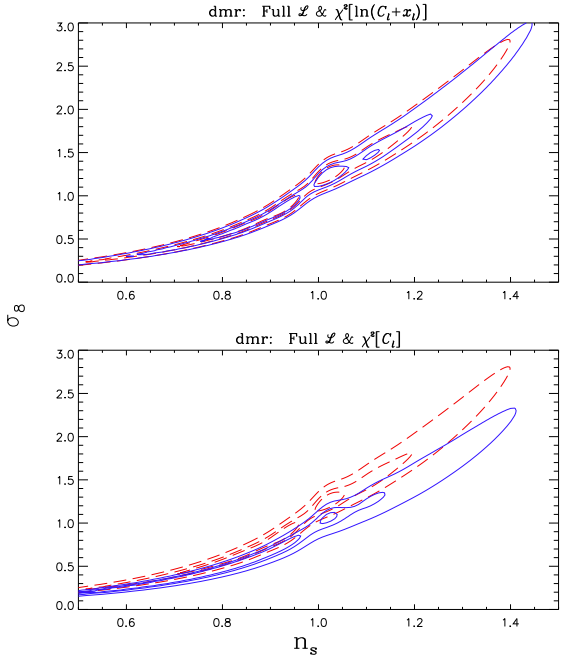}
    \caption{Exact and approximate likelihood contours for COBE/DMR,
    for the cosmological parameters $n_s$ and $\sigma_8$ (with otherwise
    standard CDM values).  Contours are for ratios of the likelihood to
    its maximum equal to $\exp{-\nu^2/2}$ with $\nu=1,2,3$. Upper panel
    is for the full likelihood (dashed) and its offset lognormal
    approximation as a 
    Gaussian in $\ln{(\C_\ell+x_\ell)}$ (solid; see text); lower panel
    shows the full likelihood and its approximation as a Gaussian in
    $\C_\ell$. \label{fig:dmrlikecontours}}
\end{figure}

Again, we see how well our offset lognormal ansatz performs; it
reproduces the peak and errors on the parameters. In particular, it
eliminates the ``cosmic bias'' discussed above, finding essentially the correct
amplitude, $\sigma_8$ for each shape probed, unlike the naive Gaussian
in $\C_\ell$, which consistently underestimates the amplitude. Of
course, far from the likelihood peak, even the offset lognormal form
misrepresents the detailed likelihood structure since no Gaussian
correctly represents the softer tails of the real distribution, which
goes asymptotically as the power law $\C_\ell^{-1/2}$; the offset
lognormal approximation is asymptotically lognormal with a much steeper
descent; the equal-variance form can in principle reproduce the asymptotic
form better.  This behavior can be important for the case of upper
limits, {\sl i.e.}, when the likelihood peak is at $\C_\ell = 0$.  We discuss
this special case in Section~\ref{sec:upperlim}.

\section{General Treatment}
\label{sec:general}
\subsection{Chopping Experiments}
\label{sec:chop}
We wish to generalize this procedure to the case of experiments
that are not capable of estimating individual multipole moments
and/or chopping experiments.
By chopping experiments, which have been the norm until very recently,
we mean those that rather than
report the temperature at various positions on the sky,
report more complicated linear combinations, with a sky
signal given by
\begin{equation}
  s_i=\int d^2{\hat x} \; H_i({\hat x}) {\Delta T\over T}({\hat x})
=\sum_{\ell m} H_{i,\ell m} a_{\ell m}
\end{equation}
for some beam and switching function $H({\hat x})$; $H_{i,\ell m}$ and
$a_{\ell m}$ are the spherical-harmonic transforms of $H$ and the
temperature, respectively. This induces a signal correlation matrix
given by
\begin{equation}
  \label{eqn:fullCT}
  C_{Tii'}=\langle s_i s_{i'} \rangle = \sum_\ell
  {\ell+1/2\over\ell(\ell+1)}W_{ii'}(\ell) \C_\ell.
\end{equation}
Here, the window function matrix, $W_{ii'}(\ell)$, generalizes the beam
$B_\ell^2$ of a mapping experiment and is given by
\begin{equation}
    \label{eqn:Wpp}
    W_{ii'}(\ell)={4\pi\over2\ell+1}\sum_m H_{i,\ell m} H^*_{i',\ell m}
\end{equation}
(this should not be confused with the ``window function,'' given by ${\bar
  W}_\ell=\sum_i W_{ii}(\ell)/N_{\rm pix}$.)  Moreover, for many
experiments, the noise structure can be considerably more complicated,
and may not be reducible to a simple noise correlation function or power
spectrum (that is, correlations in the noise may not just be functions
of the distance between points); instead, we may have to specify a
general noise matrix
\begin{equation}
  \label{eqn:CN}
  C_{Nii'}=\langle n_i n_{i'} \rangle
\end{equation}

How can we generalize our previous procedure to account for this more
complicated correlation structure?  We will take the general offset
lognormal form of the likelihood, a Gaussian in $Z_\ell=\ln(\C_\ell +
x_\ell)$, as our guide.  We have already noted that in the case of
incomplete sky coverage, or inhomogeneous noise, Eq.~\ref{eqn:dotrick}
has no solution.  Thus we are only searching for a reasonable ansatz to
try for $x_\ell$.

We begin by noting that $x_\ell$ represents the noise contribution
to the error.  For the full-sky case the ratio $\C_\ell/x_\ell$ is the ratio
of the signal contribution to the error to the
noise contribution to the error:
\begin{equation}
\label{eqn:cloverxl}
\C_\ell/x_\ell = {(F^{-1/2}_{ll})_{\rm signal} \over
(F^{-1/2}_{ll})_{\rm noise}}
\end{equation}
since $(F^{-1/2}_{ll})_{\rm signal}=\sqrt{2/(2\ell+1)}\C_\ell$ and
$(F^{-1/2}_{ll})_{\rm noise}=\sqrt{2/(2\ell+1)}\calN_\ell/B_\ell^2=
\sqrt{2/(2\ell+1)}x_\ell$.  Writing $\C_\ell/x_\ell$ in
Eq.~\ref{eqn:cloverxl} in terms of Fisher matrices allows us
to generalize to arbitrary experiments.

Before writing down the general procedure, we must introduce
a little more notation.  Instead of estimating every $\C_\ell$,
we estimate the binned power spectrum $\C_B \equiv \sum_{\ell \in B}
\C_\ell/\sum_{\ell \in B} 1 $, in bins labeled by $B$.
Let $C_{T,B}$ be the contribution to the signal covariance matrix
from bin $B$, {\sl i.e.}, $C_{T,Bij} = \sum_{\ell \in B} \C_B W_{ij}(\ell)/\ell$.
The Fisher matrix for $\C_B$ (whose inverse gives the covariance
matrix for the uncertainty in $\C_B$) is given by
\begin{equation}
F^{(\C)}_{BB'} = {\rm Tr}\left(C^{-1} C_{T,B} C^{-1} C_{T,B'}\right)/(\C_B\C_{B'})
\end{equation}
where $C$ is the total covariance matrix, $C=C_T+C_N$.  Of course, in
the limit of no noise, $C=C_T$ and in the limit of no signal, $C=C_N$.
Thus Eq.~\ref{eqn:cloverxl} generalizes to
\begin{equation}
  \label{eqn:getxb}
  \C_B/x_B = \sqrt{
    {\rm Tr}\left(C_N^{-1} C_{T,B} C_N^{-1} C_{T,B}\right)
    \over
    {\rm Tr}\left(C_T^{-1} C_{T,B} C_T^{-1} C_{T,B}\right)}.
\end{equation}

Evaluation of the denominator of Eq.~\ref{eqn:getxb} is sometimes
difficult, as practical shortcuts in the calculation of the window
function matrix may make it singular or give it negative eigenvalues.
To avoid this calculation, we sometimes generalize the expression for
$x_\ell$ by noting that for a homogeneously sampled, full-sky map:
\begin{equation}
(F_{\ell \ell}^{-1/2})_{\rm signal}=F_{\ell \ell}^{-1/2} - (F_{\ell
  \ell}^{-1/2})_{\rm noise}
\end{equation}
 and therefore replace
Eq.~\ref{eqn:getxb} with
\begin{equation}
  \label{eqn:getxb2}
  \C_B/x_B = \sqrt{
      {\rm Tr}\left(C_N^{-1} C_{T,B} C_N^{-1} C_{T,B}\right)
      \over
      {\rm Tr}\left(C^{-1} C_{T,B} C^{-1} C_{T,B}\right)} - 1.
\end{equation}
(For the all-sky, uniform noise case, $x_B$ thus defined will be
independent of $\C_B$; in a realistic experiment this
will no longer hold. In practice, we expect that the correlation matrix
at the likelihood peak would give the best value for $x_B$.)

Alternatively, we sometimes use Eq.~\ref{eqn:getxb} but make
use of the approximation
\begin{equation}
{\rm Tr}\left(C_T^{-1} C_{T,B} C_T^{-1} C_{T,B}\right) =
\sum_{\ell \in B} (2\ell+1)f_{\rm sky}
\end{equation}
which is exact for maps in the limit $f_{\rm sky} \rightarrow 1$.

To summarize, our offset lognormal ansatz is to take $Z_B=\ln (\C_B +
x_B)$ as Gaussian distributed, with $x_B$ calculated from
Eq.~\ref{eqn:getxb2} and covariance matrix given by the inverse of
\begin{equation}
F^{(Z)}_{BB'} = Z_B Z_{B'} F^{(\C)}_{BB'} \ \ {\rm no} \ {\rm sum}.
\end{equation}
Alternately, we can use these same quantities in the equal-variance
form of Eqs.~\ref{eqn:equalindepmode}--\ref{eqn:chi2approxparams}.

\subsection{Bandpowers}
Most observational power spectrum constraints to date are reported
as ``band-powers'' rather than as estimates of the power in a
power-spectrum bin, as we have been assuming above.  These band-powers
are the result of assuming a given shape for the power spectrum
and then using one particular modulation of the data to determine
the amplitude.  With $\C_\ell$ thus fixed,
the band-power, $\C_{BP}$, is given by
\begin{equation}\label{eq:bandpowerdef}
\C_{BP} = {\sum_\ell {\ell+1/2\over\ell(\ell+1)} {\bar W}_\ell \C_\ell\over
\sum_\ell {\ell+1/2\over\ell(\ell+1)}{\bar W}_\ell} \simeq
 {\sum_\ell  {\bar W}_\ell \C_\ell/\ell\over
\sum_\ell {\bar W}_\ell/\ell} \, ,
\end{equation}
with ${\bar W}_\ell$ given by the trace of the window function matrix.  To
find $x_{BP}$, replace $C_{T,B}$ with $C_T$ in
Eq.~\ref{eqn:getxb2}.

Note that in order to compare this observationally-determined number,
$\C_{\rm BP}$, with other theories, specified by different $\C_l$s with
different shapes and amplitudes, we need a means for calculating the
expected value of $\C_{\rm BP}$, $\langle \C_{\rm BP} \rangle$, given an
arbiratry $\C_l$.  It has been often assumed that $\langle \C_{\rm
  BP}\rangle$ for arbitrary $\C_l$ is also given by the right-hand side
of Eq.~\ref{eq:bandpowerdef}.  However, this is only strictly true in
the case of a diagonal signal covariance matrix.  The generalization to
the case of non-diagonal signal matrices is discussed in \citet{bjkpspec}
and \citet{Knox99}.

\subsection{Linear Combinations}

Nothing we have derived so far restricts us to the likelihood as a
function of $\C_\ell$ {\em per se}; any other measure of amplitude will
also have a likelihood in this form. That is, we write a general
amplitude as
\begin{equation}
  \sigma_{i}^2 = \sum_\ell f_i(\ell) {\cal C}_\ell
\end{equation}
for some arbitrary filter or filters, $f_{i}(\ell)$, and ${\cal
  C}_\ell=\ell(\ell+1)C_\ell/(2\pi)$ as usual. This filter
could be, for example, one designed to make the uncertainties in
the $\sigma_{i}^2$ uncorrelated, as in the following section.

What is the likelihood for this amplitude, rather than $\C_\ell$ at a
single $\ell$? We first change variables from $\C_\ell$ to $\sigma_i$ as
in Eq.~\ref{eqn:changevar}. If we choose window functions that do not
overlap in $\ell$, the inverse Fisher matrix then becomes
\begin{equation}
  F^{-1}_{ii'} = \sum_{\ell\ell'}  f_i(\ell) F^{-1}_{\ell\ell'} f_{i'}(\ell').
\end{equation}
If the original ($\C_\ell$) Fisher matrix has the simple form of
Eq.~\ref{eqn:fullcurv}, then we see that we can just filter the
individual
terms in any of the ensuing equations with the same $f_i$ and our ansatz
will still hold. Explicitly, we would expect the variable
\begin{equation}
\label{eqn:bandxb}
  Z_i = \ln\left[\sigma^2_i + \sum_\ell f_i(\ell)x_\ell\right]
\end{equation}
to be distributed as a Gaussian for the offset lognormal form; for the
equal-variance form the generalization is clear.

\subsection{Orthogonal Bandpowers}
\label{sec:ortho}
We can apply this to a particularly useful set of linear combinations,
the so-called orthogonal bandpowers
\citep{bjkpspec,tegmark,hamilton,hamilton97b,TegHam}. If we have a set
of spectral measurements ${\cal C}_B$ in bands $B$ with a weight matrix
$M_{BB'} = F_{BB'} = \langle \delta{\cal C}_B\delta{\cal C}_{B'}
\rangle$, we can form a new set of measurements which have a diagonal
error matrix by applying a transformation like $D_B = M^{1/2}_{BB'}
{\cal C}_{B'} = F^{1/2}_{BB'} {\cal C}_{B'}$. The $1/2$ power represents
any matrix such as the Cholesky decomposition or Hermitian square root
which satisfies $A^{1/2} (A^{1/2})^T = A$.  These linear combinations
will have the property that $\langle\delta{D}_B\delta{D}_{B'}
\rangle=\delta_{BB'}$ (note the similarity to the calculations of
Section~\ref{sec:solution}). For calculating the ``naive'' quantity,
\begin{eqnarray}
  \chi^2 &=& \sum_{BB'}({\cal C}_B-{\widehat{\cal C}}_B) M_{BB'}
  ({\cal  C}_{B'}-{\widehat{\cal C}}_{B'})\\
  &=& \sum_{B} (D_B-{\widehat D}_B)^2,
\end{eqnarray}
(hats here refer to observed quantities) these orthogonalized bands
don't change the results. However, because the error and Fisher matrices
are both diagonal our likelihood ansatz can be applied very cleanly,
since the off-diagonal correlations are zero and so we might expect them
to represent the exact shape of the likelihood around the peak more
accurately. Now, we will take the quantity $\zeta_B=\ln(D_B+y_B)$ to be
distributed as a Gaussian with correlation matrix
$\langle\delta\zeta_B\delta\zeta_{B'}\rangle=(D_B+y_B)^{-2}\delta_{BB'}$.
This again involves the further approximation that the correlation
structure far from the peak remains given by that near the peak (encoded 
in the curvature matrix of the distribution at the peak).

From the previous subsection, we further know that if we have a set of
quantities $x_B$ appropriate for approximating the ${\cal C}_B$
likelihood (as in Eq.~\ref{eqn:getxb2}), then we should be able to set
$y_B=F^{1/2}_{BB'}x_{B'}$.  (Note that we can also use these quantities
in the equal-variance approximation, which does not otherwise have a
simple multivariate generalization.)

We use these orthogonalized bandpower results
for the cosmological parameter estimates using the SK data in the
following sections.

\section{Application to Saskatoon}\label{sec:SK}
We apply this ansatz to the Saskatoon experiment, perhaps the apotheosis
of a chopping experiment.  The Saskatoon data are reported as
complicated chopping patterns ({\sl i.e.}, beam patterns, $H$, above) in
a disk of radius about $8^\circ$ around the North Celestial Pole. The
data were taken over 1993-1995 (although we only use the 1994-1995 data)
at an angular resolution of $1.0$--$0.5^\circ$ FWHM at approximately
30~GHz and 40~GHz. More details can be found in \citet{nett95} and \citet{woll95}.
The combination of the beam size, chopping pattern, and sky coverage
mean that Saskatoon is sensitive to the power spectrum over the range
$\ell=50$--$350$.  The Saskatoon dataset is calibrated by observations
of supernova remnant, Cassiopeia--A.\ Leitch and collaborators
\citep{ELthesis} have recently measured the flux and find that the
remnant is 5\% brighter than the previous best determination.  We have
renormalized the Saskatoon data accordingly.

\begin{figure}
  \plotone{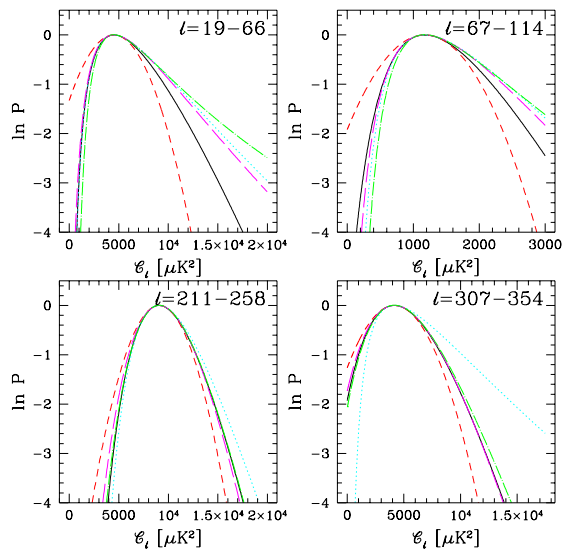}
  \caption{Full and approximate Saskatoon likelihoods. As in
    Fig.~\ref{fig:dmrlike}. The solid (black) line is the full likelihood
    calculated exactly. The short-dashed (red) line is the Gaussian
    approximation about the peak. The dotted (cyan) line is a Gaussian in
    $\ln{\C_\ell}$; the  dashed (magenta) line is a Gaussian in
    $\ln{(\C_\ell+x_\ell)}$, as discussed in the text. The 
    dot-dashed (green) line is the equal-variance approximation.
    \label{fig:sklike}}
\end{figure}

We calculated $\C_\ell$ for this dataset in \citet{bjkpspec}. We combine
these results with the data's noise matrix to calculate the appropriate
correlation matrixes (in this case, the full curvature matrix) for
Saskatoon and hence the appropriate $x_B$ (Eq.~\ref{eqn:getxb2}) and
thus our approximations to the full likelihood.  In
Figure~\ref{fig:sklike}, we show the full likelihood, the naive Gaussian
approximation, and our present offset lognormal and equal-variance
forms.  Again, both approximations reproduce the features of the
likelihood function reasonably well, even into the tails of the
distribution, certainly better than the Gaussian approximation.  They
seem to do considerably better in the higher-$\ell$ bands; even in the
lower $\ell$ bands, however, the approximations result in a {\em
  wider}\/ distribution which is preferable to the narrower Gaussian and
its resultant strong bias.  Moreover, we have found that we are able to
reproduce the shape of the true likelihood essentially perfectly down to
better than ``three sigma'' if we simply {\em fit}\/ for the $x_B$ (but
of course this can only be done when we have already calculated the full
likelihood---precisely what we are trying to avoid!). For existing
likelihood calculations, this method can provide better results without
any new calculations (see Appendix~\ref{app:recipe} for our
recommendations for the reporting of CMB bandpower results for extant,
ongoing, and future experiments).

\begin{figure}
  \plotone{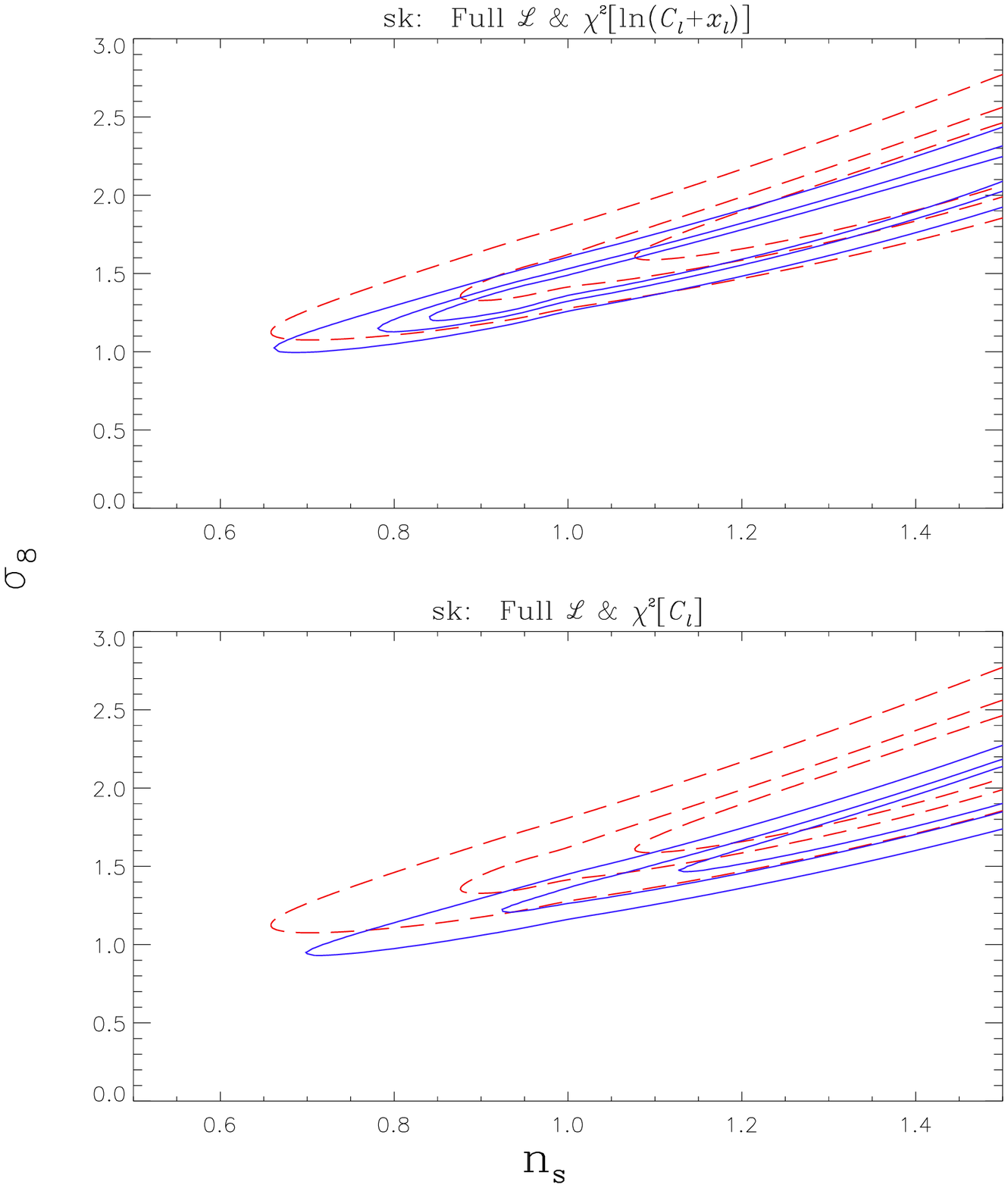}
    \caption{Likelihood contours for the Saskatoon experiment alone, as
    in Figure~\ref{fig:dmrlikecontours}, but using the ``orthogonalized
    bands'' of Sec.~\ref{sec:ortho}. \label{fig:skcontours}}
\end{figure}

We also show the usefulness of the offset lognormal form in cosmological
parameter determination in Figure~\ref{fig:skcontours}, for which we
use the orthogonalized bandpowers discussed above. As expected, it
reproduces the overall shape of the likelihood function quite well,
although it does better for a fixed shape ($n_s$ in this case).

We have found that the shape of the power spectrum used with each bin
of $\ell$ can have an impact on the likelihood function
evaluated using this ansatz. Similarly, a finer binning in
$\ell$ will reproduce the full likelihood more accurately.
Although the maximum-likelihood amplitude
at a fixed shape ($n_s$) does not significantly depend on binning or
shape, the shape of the likelihood function along the
maximum-likelihood ridge changes with finer binning and with the assumed
spectral shape.

As an aside, we mention several complications that we have noted in the
analysis of the Saskatoon data. Because of the complexity of the Saskatoon chopping
strategy, we have found that the signal correlation matrix, $C_T$ is
not numerically positive definite; removing the negative eigenvalues can
change the value of $\C_B$ by as much as 5\% in some bins. This should
be taken as an estimate of the accuracy of our spectral determinations
due to these numerical errors.

We have also found that the Fisher matrix, which we usually use as an
estimate of the (inverse) error matrix for the parameters, can differ
significantly from the true curvature matrix. This difference can be
especially marked in low-$\ell$ bins for which the sample and/or cosmic
variance can be considerable, potentially resulting in large
fluctuations in this error estimate as well. In the Saskatoon plots
here, we use the actual curvature matrix in place of the Fisher matrix.
In a forthcoming paper \citep{knoxjaffequadest}, we will address these
and other issues of implementation of the quadratic estimator for
$\C_\ell$.

As we will see in the following, these concerns become less important
when combining Saskatoon with COBE/DMR, since the results are mostly
dependent on the broad-band power probed by each experiment. Moreover,
we expect that these difficulties are considerably more likely in the
case of chopping experiments, for which our expression for $x_B$,
Eq.~\ref{eqn:getxb} is somewhat ad hoc. Most future CMB results will be
for ``total-power'' ({\sl i.e.}, mapping) experiments, and the
satellites MAP and Planck will be (nearly) all-sky, like COBE/DMR, for
which the offset lognormal form has proven most excellent. In any case,
even with present-day data, our ansatz provides a far better
approximation to the full likelihood than a simple Gaussian in $\C_\ell$
as was used for some global analyses of current CMB data such as
\citet{lineweaver97}; \citet{lineweaver98a}; \citet{lineweaver98b};\citet{lineweaver98c};\citet{lineweaverBarbosa97};\citet{hancockrocha}).

\section{Application to OVRO, SP and SuZIE}
\label{sec:upperlim}

\begin{figure}
  \plotone{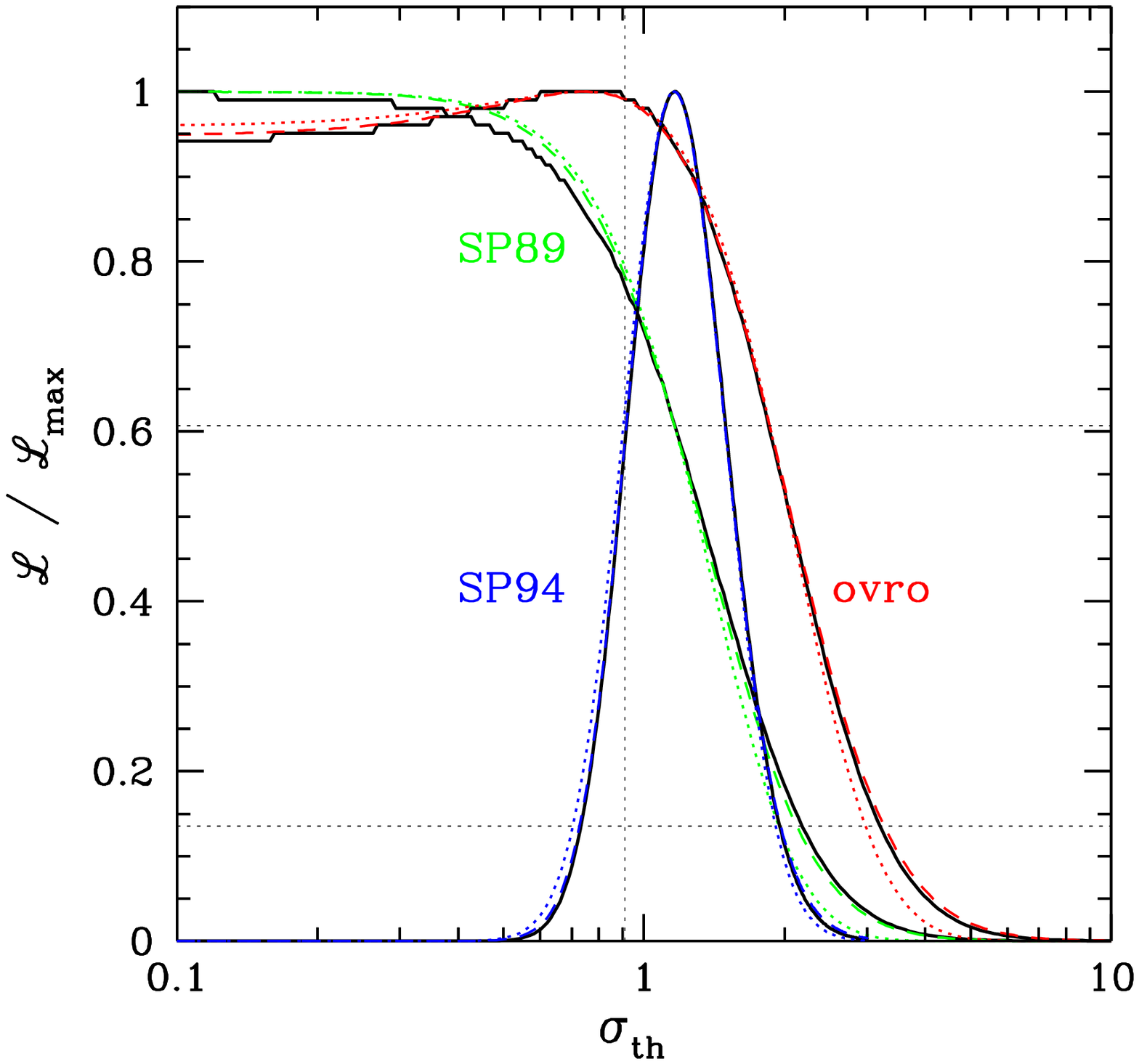}\\ \plotone{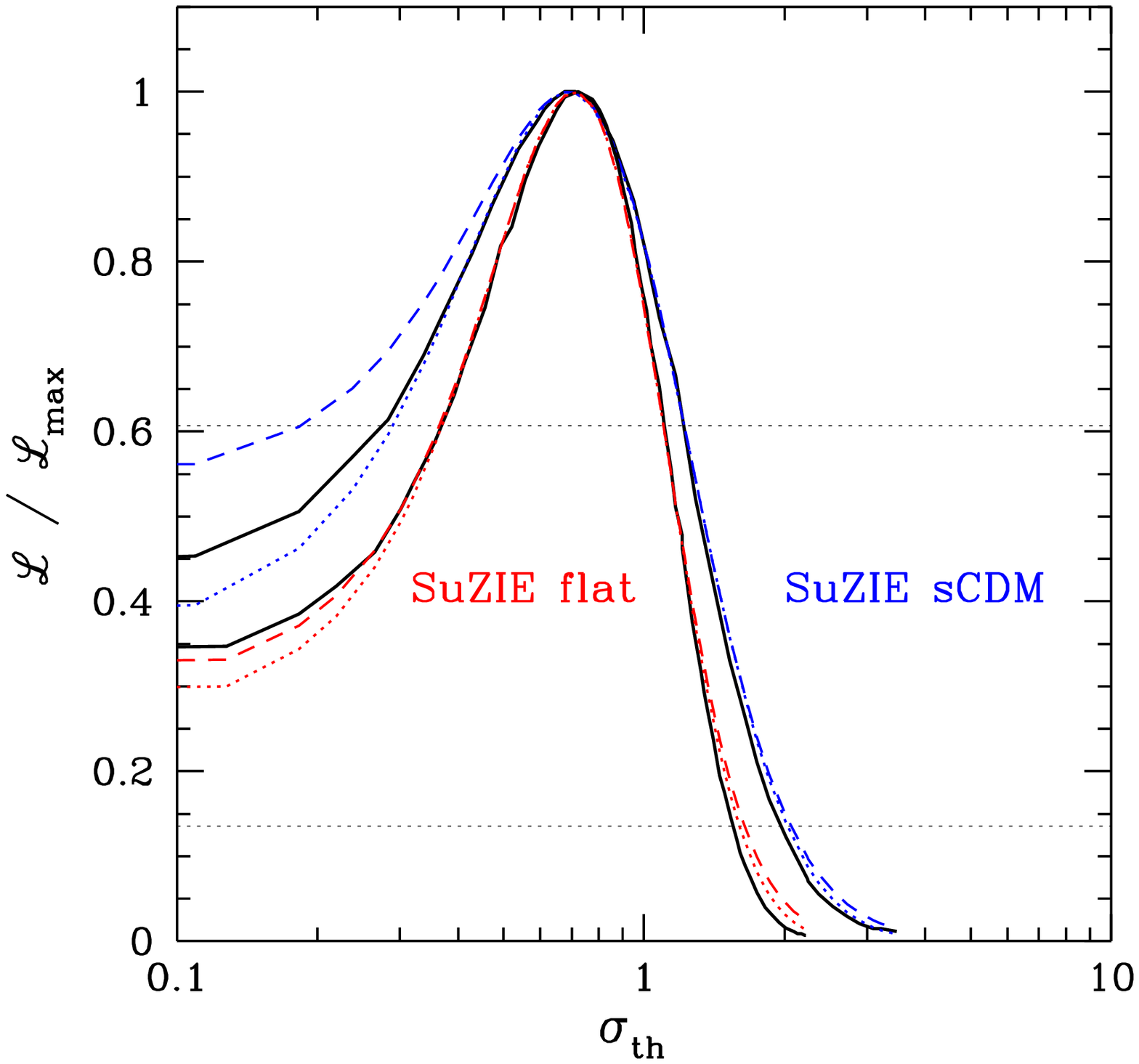}
  \caption{Upper Panel: Likelihood curves for two upper limit cases,
    OVRO and SP89, and SP94 which had a reasonably well determined
    bandpower amplitude. Lower Panel: Likelihood curves for SuZIE, using
    two different models for $\C_\ell$, sCDM and a flat bandpower. Solid
    (black) lines are the full likelihoods.  The equal-variance
    approximation (dashed curve) does extremely well and the
    offset-lognormal (dotted curves) also does well in treating these
    cases. The measure of the amplitude $\sigma_{th}$ is proportional to
    ${\cal C}_B^{1/2}$. In the top panel, $\sigma_{\rm th}=\sigma_8$ for
    the untilted sCDM model; in the bottom panel, it is in units of
    $10^5 \Delta T/T$.
    \label{fig:ovrolike}}
\end{figure}

One of the problems we hoped to solve with better approximations to
the likelihood functions than Gaussian was how to treat the valuable
data with upper limits or very weak detections. In particular, the
data from OVRO \citep{OVRO} and SuZIE \citep{SuZIE} is useful for
constraining open universe models with power spectra that do not fall
off rapidly enough at high $\ell$.  Although the Gaussian form
does not work well here, the offset lognormal does much better
and the form of Section~\ref{sec:indmode} works very well, as is
shown for OVRO and SP in the top panel of Fig.~\ref{fig:ovrolike}.

The likelihood for the SuZIE results is also shown. The authors
\citep{SuZIE} plotted the likelihood for the amplitude for several
different models, which we have fit from the published figure. Although
reported as an upper limit, the likelihood is peaked at positive power,
but zero power is only rejected at $\sim1\sigma$.  We note as an aside
that a simple flat bandpower ($\C_\ell={\rm const}$) is not quite
sufficient to contain all of the information in the SuZIE data: the
likelihood function changes slightly for models with different shapes,
defined as in Eq.~\ref{eq:bandpowerdef}---most of the physically
motivated models ({\sl e.g.}, sCDM, $\Lambda$CDM, etc.) have roughly the same
bandpower curves, but a flat bandpower gives a slightly different one as
shown in the figure, and extreme open models are more similar to the
flat-bandpower case than to sCDM.  We also note that our equal-variance
approximation performs slightly better for the flat bandpower, while the
sCDM model is fit better by the offset lognormal. In any case, we again
find that in all of these cases our approximations fit the likelihoods
much better than any naive Gaussian approach would.

\section{ Results: COBE/DMR + Saskatoon}
\label{sec:results}
\begin{figure}
  \plotone{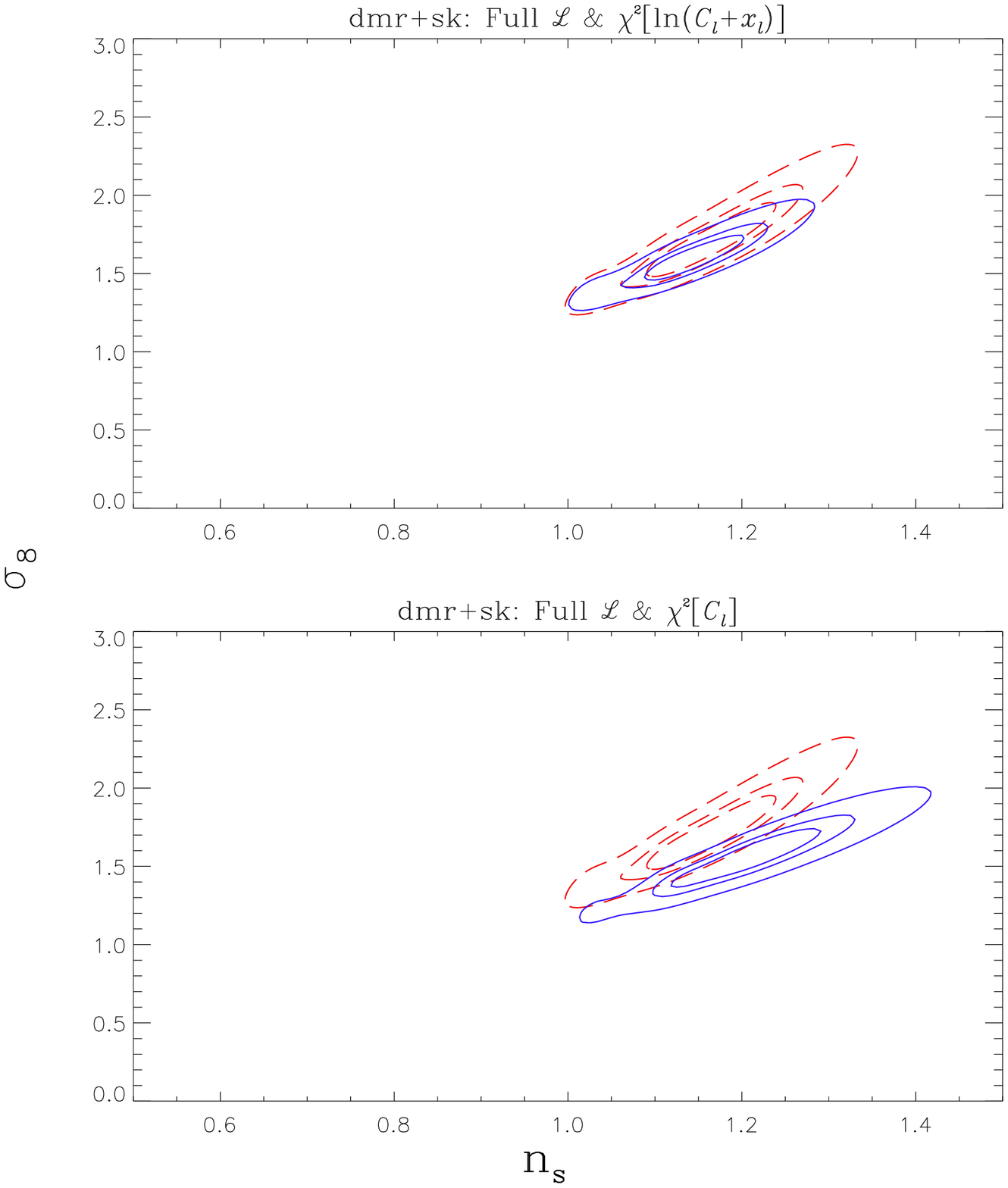}
    \caption{Likelihood contours for COBE/DMR and Saskatoon combined.
    As in Figure~\ref{fig:dmrlikecontours}, but combining likelihoods
    from COBE/DMR and Saskatoon. For the Saskatoon calculation, we used
    the ``orthogonalized bands'' of Sec.~\ref{sec:ortho}.
    \label{fig:dmrskcontours}}
\end{figure}

As a further example and test of these methods, we can combine the
results from Saskatoon and COBE/DMR in order to determine cosmological
parameters. For this example, we use the orthogonal linear combinations
as described in the previous section.  In Figure~\ref{fig:dmrskcontours}
we show the likelihood contours for standard CDM, varying the scalar
slope $n_s$ and amplitude $\sigma_8$.  As before, we see that the naive
$\chi^2$ procedure is biased toward low amplitudes at fixed shape
($n_s$), but that our new approximation recovers the peak quite well.
The full likelihood gives a global maximum at $(n_s,
\sigma_8)=(1.15,1.67)$, and our approximation at $(1.13,1.58)$, while
the naive $\chi^2$ finds it at $(1.21,1.55)$, outside even the
three-sigma contours for the full likelihood. We can also marginalize
over either parameter, in which case the full likelihood gives
$n_s=1.17^{+0.08}_{-0.07}$, $\sigma_8=1.68^{+0.26}_{-0.21}$; our ansatz
gives $n_s=1.14^{+0.07}_{-0.05}$, $\sigma_8=1.60\pm0.15$; and the naive
$\chi^2$ gives $n_s=1.21^{+0.08}_{-0.09}$,
$\sigma_8=1.55^{+0.18}_{-0.20}$.
(Note that even with the naive $\chi^2$ we marginalize by explicit
integration, since the shape of the likelihood in parameter space is
non-Gaussian in all cases.)

\section{Parameter Estimation}
\label{sec:bands}
Above, we have discussed many different approximations to the likelihood
${\cal L}$.  Here we discuss finding the parameters that maximize this
likelihood (minimize the $\chi^2\equiv-2\ln{\cal L}$).  We then apply
our methods to estimating the power in discrete bins of $\ell$.  This
application provides another demonstration of the importance of using a
better approximation to the likelihood than a Gaussian.

The likelihood functions above depend on ${\C_\ell}$ which may
in turn depend on other parameters, $a_p$, which are, {\sl e.g.},
the physical parameters of a theory.  If we write the parameters as
$a_p +\delta a_p$ we can find the correction, $\delta a_p$,
that minimizes $\chi^2$ by solving
\begin{equation}
  \label{eqn:quadest}
  \delta a_p = -\frac{1}{2}{\cal F}^{-1}_{pp'} { \partial \chi^2 \over
    \partial a_{p'}},
\end{equation}
where
\begin{equation}\label{eqn:chi2curv}
{\cal F}_{pp'} \equiv \frac{1}{2}{ \partial^2 \chi^2 \over
\partial a_{p}\partial a_{p'} }
\end{equation}
is the curvature matrix for the parameters $a_p$.  If the $\chi^2$ were
quadratic ({\sl i.e.}, Gaussian ${\cal L}$) then Eq.~\ref{eqn:quadest} would
be exact.  Otherwise, in most cases, near enough to its minimum,
$\chi^2$ is approximately quadratic and an iterative application of
Eq.~\ref{eqn:quadest} converges quite rapidly.  The covariance matrix
for the uncertainty in the parameters is given by $\langle \delta a_p
\delta a_{p'} \rangle = {\cal F}^{-1}_{p p'}$. This is just an approximation to
the Newton-Raphson technique for finding the root of $\partial{\cal
  L}/\partial a_p = 0$; a similar techniqe is used in 
quadratic estimation of $\C_\ell$
\citep{tegmark,bjkpspec,OhSpergelHinshaw}.

As our worked example here, we parameterize the power spectrum
by the power in $B=1$ to $11$ bins, $\C_B$.  Within each of the bins,
we assume $\C_\ell=\C_B$ to be independent of $\ell$.
We have chosen the offset lognormal approximation.  The $\chi^2$
completely describes the model:
\begin{eqnarray}
\label{eqn:chisq}
\chi^2 & =& \sum_{i,j}\left(Z_i^{\rm t}-Z_i^{\rm d}\right)M_{ij}^Z
\left(Z_j^{\rm t}-Z_j^{\rm d}\right) + \chi^2_{\rm cal}; \\
\chi^2_{\rm cal} &\equiv & \sum_\alpha {(u_\alpha-1)^2
\over \sigma_\alpha^2}; \\
Z_i^{\rm d} & \equiv & \ln(D_i+x_i); \\
Z_i^{\rm t} &\equiv&
\ln\left(\sum_B u_{\alpha(i)}f_{iB}\C_B + x_i\right); \\
M_{ij}^Z & \equiv & M_{ij}\left(D_i+x_i\right)\left(D_j+x_j\right)\quad
\mbox{no sum};
\end{eqnarray}
where $M_{ij}$ is the weight matrix for the band powers
$D_i$\footnote[2]{In most cases, it is more precisely an {\it estimate} of
the
weight matrix based on, {\sl e.g.}, 68\% confidence upper and lower limits.
For more details, see Appendix~\ref{app:Bandpowers}.}.  We have
modeled the signal contribution to the data, $D_i$, as an average over
the power spectrum, $\sum_B f_{iB}\C_B$, times a calibration parameter,
$u_{\alpha(i)}$.  For simplicity, we take the prior probability
distribution for this parameter to be normally distributed.  Since the
datasets have already been calibrated, the mean of this distribution
is at $u_{\alpha}=1$.  The calibration parameter index, $\alpha$, is a
function of $i$ since different power spectrum constraints from the
same dataset all share the same calibration uncertainty.  We solve
simultaneously for the $u_{\alpha}$ and $\C_B$; {\sl i.e.}, together they
form the set of parameters, $a_p$, in Eq.~\ref{eqn:quadest}.  For
those experiments reported as band-powers together with the trace of
the window function, $W^i_\ell$, the filter is taken to be
\begin{equation}
\label{eqn:win2filt}
f_{iB}={\sum_{\ell \in B} W^i_\ell/\ell \over \sum_\ell W^i_\ell/\ell}.
\end{equation}
Note that this is an approximation which neglects some effects
of off-diagonal signal correlations \citep{Knox99}.

For Saskatoon and COBE/DMR, our $D_i$ are themselves estimates of the
power in bands.  For these cases the above equation applies, but with
$W_\ell/\ell$ set to a constant within the estimated band and zero
outside.  The estimated bands have different $\ell$ ranges than the
target bands.

Instead of the curvature matrix of Eq.~\ref{eqn:chi2curv} we use an
approximation to it that ignores a logarithmic term.  Including this
term can cause the curvature matrix to be non-positive definite as the
iteration proceeds.  The approximation has no influence on our
determination of the best fit power spectrum, but does affect the error
bars.  We have found that the effect is quite small.

We now proceed to find the best-fit power spectrum given different
assumptions about the value of $x_i$, binnings of the power spectrum and
editings of the data.  See Appendix~\ref{app:Bandpowers} for a
tabulation of the bandpower data we are using.

\begin{figure}
  \plotone{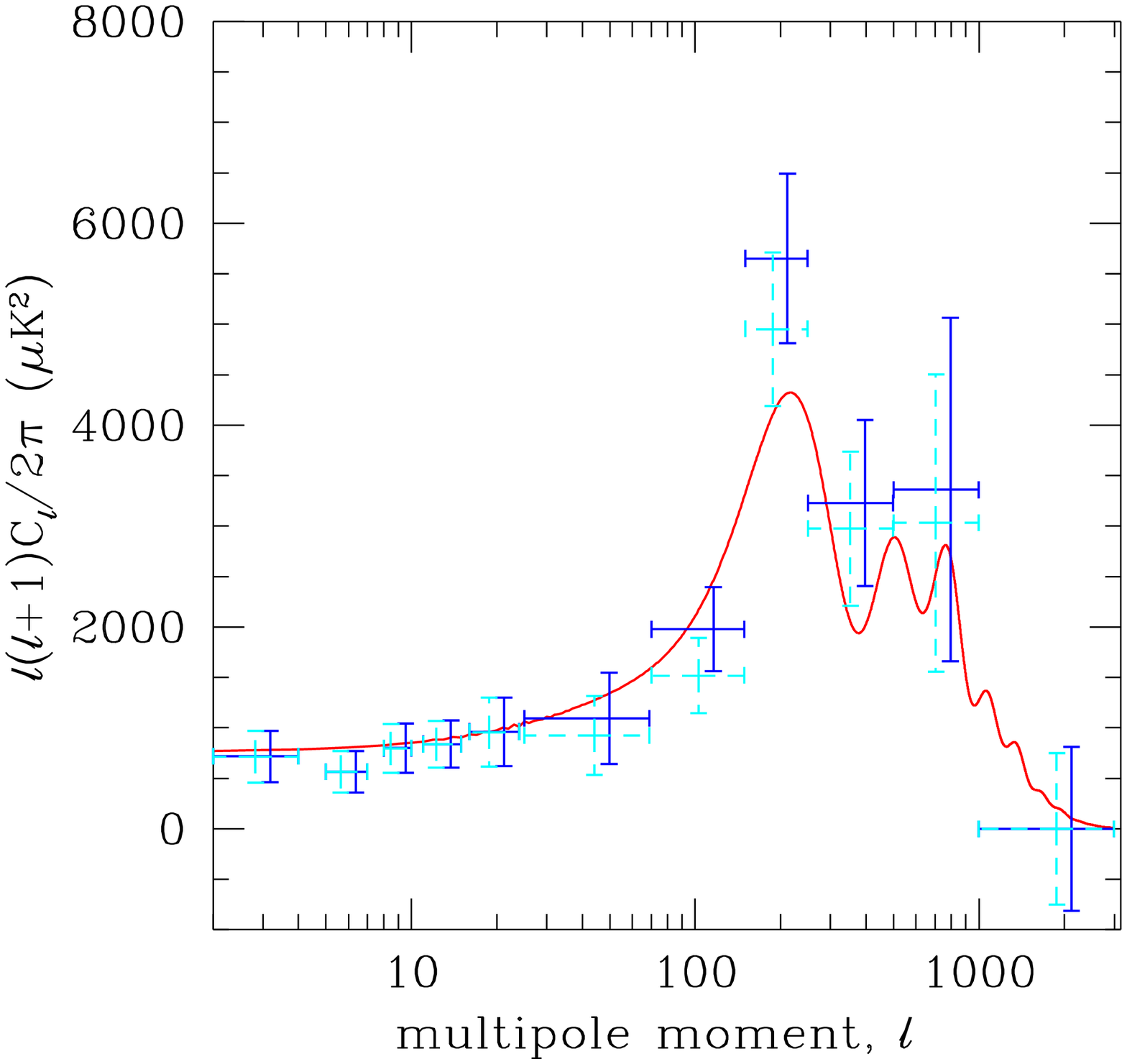}
  \caption{Power
    spectra that minimize the $\chi^2$ in Eq.~\ref{eqn:chisq}.  The
    solid (dashed) error bars assume $x = 0$ ($x=\infty$) for those
    datasets with no determination of $x$; the two sets have been offset 
    slightly for display purposes. Solid curve is standard CDM.
\label{fig:lnclvscl}}
\end{figure}

We have determined the $x_i$ only for COBE/DMR, Saskatoon, SP89, OVRO7,
SuZIE and TOCO. (Although not included in Table \ref{tab:Bandpowers},
the highly constraining measurement near $l=400$ by TOCO98 \citep{mat98} is also
well-fit by the offset lognormal form.)  To test the sensitivity to the
unknown $x_i$s we found the minimum-$\chi^2$ power spectrum assuming the
two extremes of $x_i=0$ (corresponding to lognormal) and $x_i=\infty$
(corresponding to Gaussian).  These two power spectra are shown in
Fig.~\ref{fig:lnclvscl}.  Note that both power spectra were derived
using our measured $x_i$ values; only the unknown $x_i$ values were
varied.  The variation in the results would be much greater if we let
these $x_i$ values be at their extremes.  In what follows the unknown
$x_i$ are set to zero.

Sometimes when the entire likelihood function is unavailable, enough
information is given to allow an estimation of the $x_i$ for
individual experiments that give single bandpower estimates. For
example, we may be given $D_m$ where the likelihood is a maximum and
$D_{\pm}$ where the likelihood has fallen by $\exp(-1/2)$, the latter
giving an estimate of asymmetric 1 sigma error bars. We can then
estimate the error $\sigma$ and the value of $x$ in the offset
lognormal form from:
\begin{eqnarray}
&& \sigma = \ln(s_{+}/s_{-}) \, , \nonumber \\
&& x = D_m\left[s_{+}s_{-}/(s_{+}-s_{-})-1\right] \label{eq:xs} \, , 
\end{eqnarray}
where 
\begin{eqnarray}
&& D_{\pm} \equiv D_m (1 \pm s_{\pm})
\end{eqnarray}
defines $s_{\pm}$. Note that $x$ is indeterminant for $s_{+}=s_{-}$,
but this Gaussian limit is approached from the large $x$ direction. It
yields an error on $D-D_m$ of $x\sigma \rightarrow D_m(s_{+}+s_{-})/2$, the required Gaussian error.

Although the determination of the effective error is quite stable as
$s_{-}\rightarrow s_{+}$, this is not so for $x$, where small errors
in $s_{-}, s_{+}$ can lead to big changes.  The $x>0$ constraint
implies $s_{-}s_{+} > (s_{+}-s_{-})$, which can still be violated
though $s_{+}> s_{-}$. Even if this does occur, the lognormal form
may still be a reasonable fit if one adjusts $x$ and $\sigma$. On the
other hand, if the probability is skewed towards small amplitudes, as
can happen when there is another signal such as a dust component that
has been marginalized, the lognormal fit can never work.

Most often, the reason we cannot get good values of $x$ from asymmetric
error bars is that the limits quoted in the literature are the Bayesian
integral 16\% and 84\% probability ones, with an assumed prior ({\sl
  e.g.}, uniform in the bandpower amplitude). We need to know the
likelihood shape to estimate $x$ from those $s_{\pm}$. (One could adopt
the lognormal and fit for these integral quantities, a step we have not
taken.) For example, We have found that adopting Eq.~(\ref{eq:xs}) leads
to negative values for $x$ for Tenerife, MAX4 and MAX5 values taken from
the literature.

Except for the cases noted, we always use $x$'s for the experiments
where we have determined them well, those given in Table C2.  When we
must resort to the Eq.~(\ref{eq:xs}) procedure to estimate them, we
usually set all $x$'s to zero if they are negative. For the positive
values, we have tested the Eq.~(\ref{eq:xs}) $x$ values with our $x=0$
results and find that it makes little difference in the compressed
bandpower results.

Our results have some sensitivity to binning, partly because the
visual representation does not describe the correlation between
bins. This is especially so for the last bin for which upper limits
play a large role. A procedure often used to control such variations
is to introduce a prior probability $P_P({\cal C}_B)= \exp[S_P({\cal
C}_B)]$ for the ${\cal C}_B$. In this section, we have implicitly
adopted a uniform prior, which is the least prejudiced one to adopt in
that only the data decides what the bandpowers should be. However, we
have also experimented with other priors, such as a Gaussian priors,
with $S_P \propto -({\cal C}_B/{\cal C}_{B*}-1)^2/(2\sigma_P^2)$ and a
``maximum entropy'' prior, 
\begin{eqnarray}
S_P &=& \alpha {\cal C}_{B*} \times\nonumber\\
&& [-({\cal C}_B/{\cal C}_{B*})  \ln({\cal C}_B/{\cal C}_{B*}) + ({\cal
C}_B/{\cal C}_{B*} -1) ].  \nonumber\\&&
\end{eqnarray}
Here ${\cal C}_{B*}$ is some target value, associated with an assumed
prior model, $\sigma_P^2$ is an adjustable variance in the Gaussian
case and $\alpha$ is an adjustable parameter in the maximum entropy
case. For example, the maximum entropy prior is designed to avoid
negative ${\cal C}_B$, relaxes to ${\cal C}_{B*}$ where there is
little data and ensures some degree of smoothness in the
determinations. However, especially in regions where the input shape
for ${\cal C}_{B*}$ is changing rapidly ({\sl e.g.}, the damping tail), this
procedure can give the wrong impression. Given the current state of
the data, we prefer to show a few binnings to demonstrate
sensitivity. When the data improves, it will be reasonable to try
other priors, for example those that exert penalties if the data does
not give continuity in $\ell$.

\begin{table}
\begin{tabular}{|r|r|r|r|r|}
\tableline
$\ell_{\rm min}$ & $\ell_{\rm max}$ &
power& standard error &correlation\\ \tableline
     2 &     4 &    721 &    255 & -0.08\\
     5 &     7 &    566 &    205 & -0.11\\
     8 &    10 &    799 &    242 & -0.08\\
    11 &    15 &    842 &    233 & -0.11\\
    16 &    39 &    907 &    313 & -0.17\\
    40 &    99 &   1027 &    331 & -0.32\\
   100 &   169 &   3205 &    610 & -0.16\\
   170 &   249 &   6216 &   1040 & -0.15\\
   250 &   399 &   2394 &   1106 & -0.62\\
   400 &   999 &   3328 &   1218 & -0.31\\
  1000 &  2999 &    0.0 &    783 & \\
\tableline
\end{tabular}
\caption{Estimated binned power spectrum.  The power and standard error
are in $\muK^2$ and are the numbers corresponding to the data points
in Fig.~\ref{fig:final}.  The correlation column lists the correlation
($\equiv {\cal F}^{-1}_{BB'}/\sqrt{{\cal F}^{-1}_{BB}{\cal
F}^{-1}_{B'B'}}$) between bin $B$
and the next highest bin, $B'=B+1$. 
\label{tab:thepowspec}}
\end{table}

The $\chi^2$ (of Eq.~\ref{eqn:chisq}) for the fit in Table 1 is
62 for 63 degrees of freedom.  Thus 
the scatter of the band powers is consistent with the size of
their error bars; it provides no evidence for contamination,
mis-estimation of error bars, or severe non-Gaussianity
in the probability distribution of the underlying signal.

\begin{figure}
  \plotone{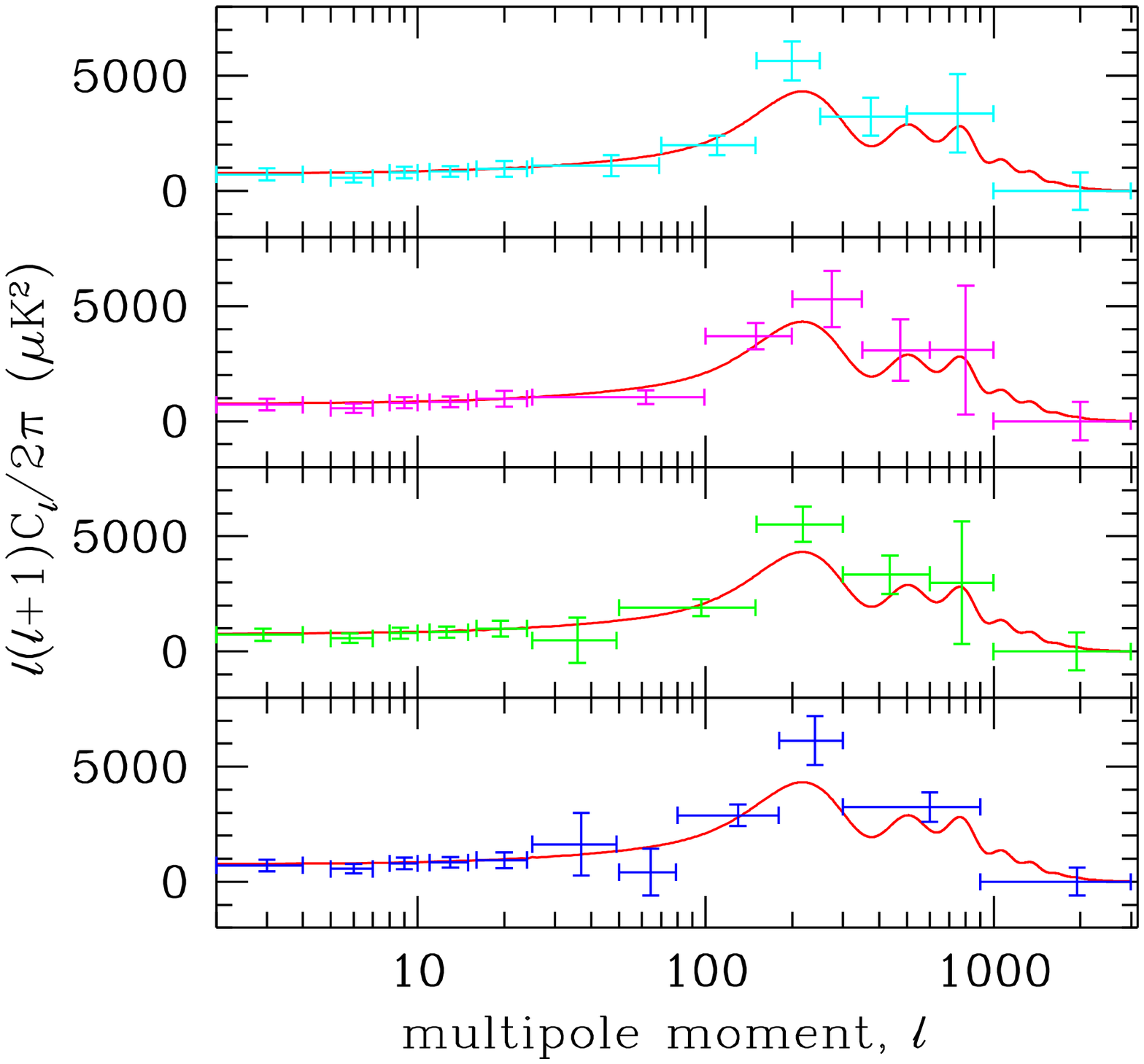}
  \caption{Power spectra that minimize the $\chi^2$ in
    Eq.~\ref{eqn:chisq} under different binnings. Solid curve is
    standard CDM.
\label{fig:binning}}
\end{figure}

In choosing a particular binning, there is a tradeoff to be
made between preserving shape information and reducing both
error bars and correlation.
From Table~\ref{tab:thepowspec} one can see the extent to which the
bins are correlated.  The
nearest-neighbor bin is by far the dominant off-diagonal term.  
For this particular binning, all
others are ten percent or less, except for the ninth bin---eleventh
bin correlation which is 0.2.  The lower $\ell$ bands have
the smallest correlations as we would expect from DMR. There
are some very strong correlations in the higher bands.  
Fortunately, from Fig.~\ref{fig:binning}
we see that some general features are quite robust under different
binnings.  Namely, the spectrum is flat out to $\ell \simeq 80$ or $100$,
there is a rise to $\ell \simeq 250$ or $350$ and then a drop in
power beyond $\ell \simeq 350$.  Although the data clearly indicate a
peak, it is difficult to locate the position to better than $\pm 70$.
In the top panel the rise to the Doppler peak has been binned more
finely than the others.  This is a particularly difficult area
to resolve with current data:  the correlation between the sixth
and seventh bins is $-0.75$.  

\begin{figure}
  \plotone{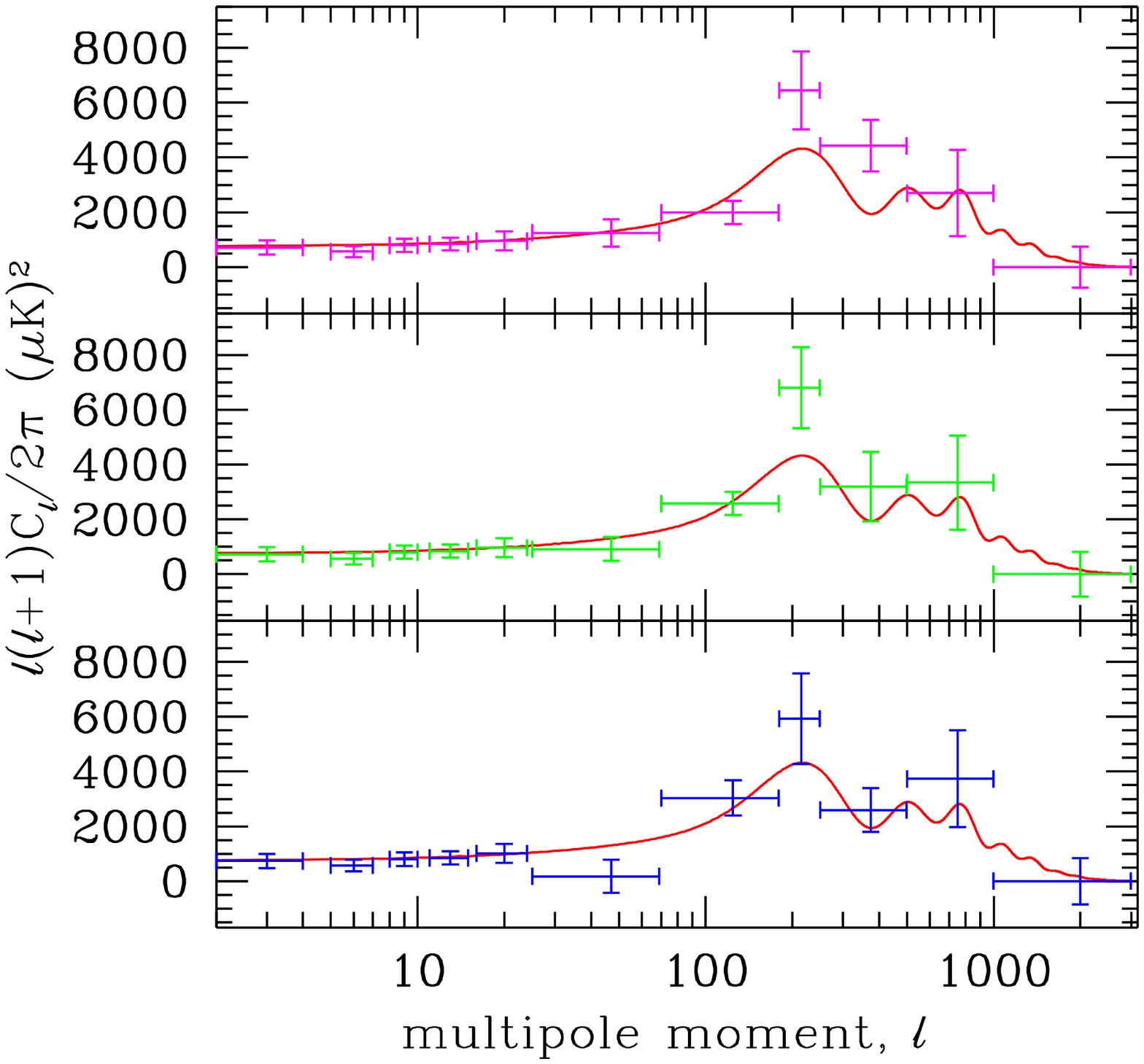}
  \caption{Power spectra that minimize the $\chi^2$ in
    Eq.~\ref{eqn:chisq} under different editings of the data.  Top
    panel: no TOCO.  Middle panel: no MSAM, bottom panel: no Saskatoon.
    Solid curve is standard CDM.
    \label{fig:editing}}
\end{figure}

We also see, from Figure~\ref{fig:editing}, that the general picture
does not depend on one single dataset---though the error bars do get
significantly larger when the Saskatoon dataset is ignored.  Also,
if we were to ignore OVRO, the highest $\ell$ bin would have 
error bars larger than the graph.  

\begin{figure}
  \plotone{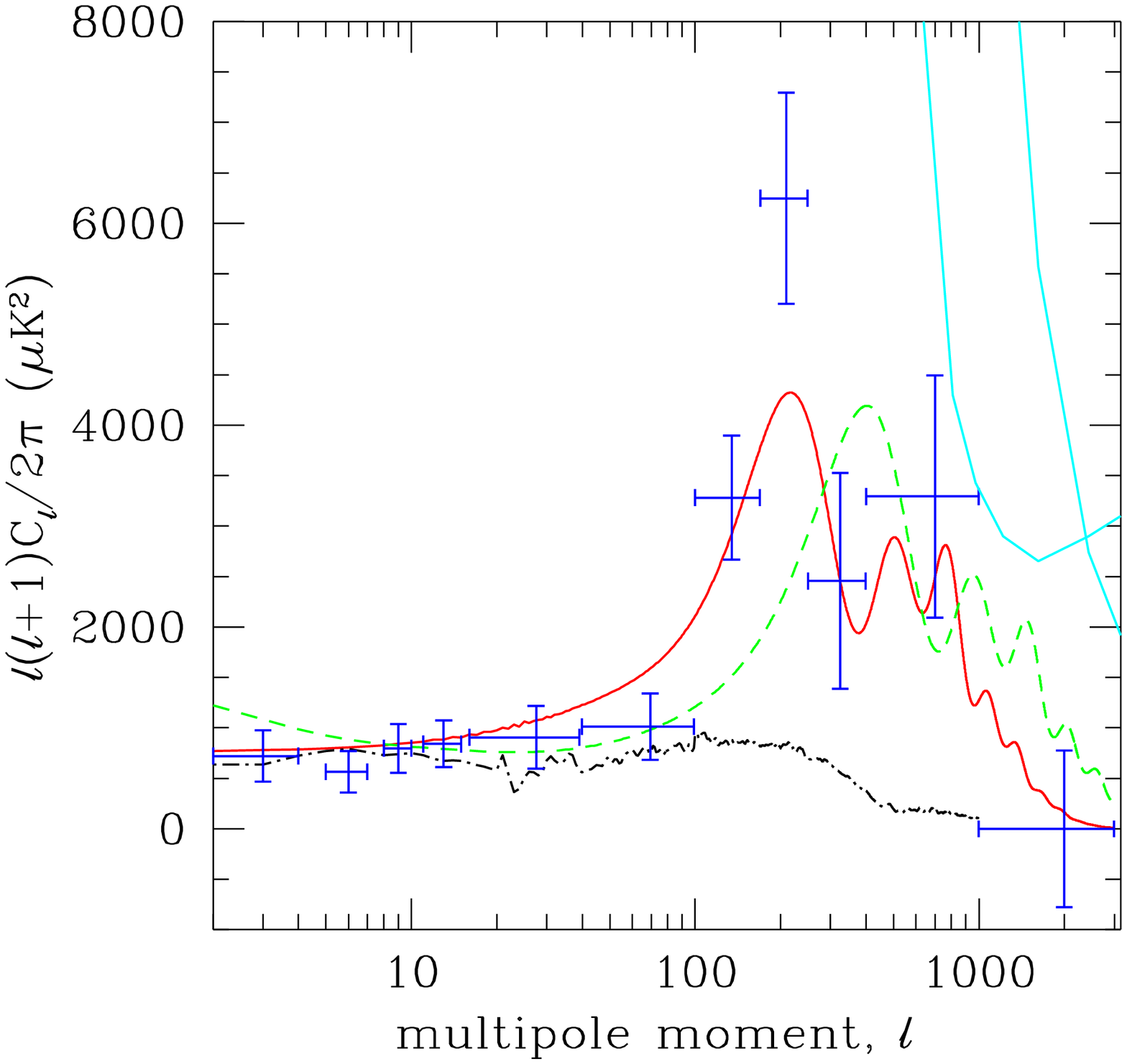} \caption{Binned power spectrum that minimizes
    the $\chi^2$ in Eq.~\ref{eqn:chisq} and which is tabulated in
    Table~\ref{tab:thepowspec}. In addition to the data presented in
    Table~\ref{tab:Bandpowers} these results also include
    TOCO98\citep{mat98}. Solid curve is standard CDM, dashed curve is an
    open CDM model with matter density one third of critical, baryon
    mass density of 0.035 times critical and a Hubble constant of
    60~km/sec/Mpc.  The dotted curve is a prediction for local cosmic
    strings \citep{allenetal}.  Curves at high $\ell$ indicate upper
    limits derived from OVRO (left) and SuZIE (right) data.
    \label{fig:final}}
\end{figure}

The upper limits from SuZIE and OVRO constrain a region of
the power spectrum otherwise unconstrained and put some
pressure on models with small-scale power, such as open models.
Due to the high interest in these limits, and the difficulty
in interpreting the error bars in these figures (once again
due to their non-Gaussianity) we have attempted to display
their constraints on the spectrum in an additional, independent
manner.  We do so by using the published
bounds on Gaussian auto-correlation functions
(GACFs).  The GACF is given by
\begin{equation}
\C_\ell = C_0 \ell^2 \theta_c^2 \exp\left(-{1\over 2}\ell^2
\theta_c^2\right)
\end{equation}
where $C_0$ is the amplitude of the real-space correlation function
at zero lag and $\theta_c$ is called the Gaussian coherence angle.
For various choices of the coherence angle, the data were used
to set limits on $C_0$.  The curves in Fig.~\ref{fig:final} trace
the peak of the GACF with $C_0$ at the 95\% confidence upper limit
as $\theta_c$ is varied.

As we have emphasized earlier with the band-powers, it can
often be misleading to interpret the covariance
matrix of the parameters (derived from the Fisher matrix) as indicating
the
68\% confidence region, since the 68\% confidence region
may extend beyond the region of validity for the quadratic
approximation to $\chi^2$.  Even in such a case though, the
quadratic procedure still may be useful just for finding
the minimum, which might be a good point to begin further investigation
of the $\chi^2$ surface without the quadratic approximation.
Non-Gaussianity can be especially severe when the parameters
are cosmological parameters.

\section{Summary and Discussion}
\label{sec:conclude}

We have argued that cosmological parameters should be constrained from
CMB datasets via an initial step of determining constraints on the power
spectrum.  These power spectrum constraints can themselves be viewed as
a compressed version of the pixelized data.  We call this process
radical compression since the resulting dataset is orders of magnitude
smaller than the original.  One must be careful in using this compressed
data to take into account the non-Gaussian nature of their probability
distributions; ignoring the non-Gaussianity while attempting to
constrain parameters results in a bias.  The offset lognormal and equal
variance approximations capture its salient characteristics.  They are
both specified by the mode, variance and the noise contribution to the
variance ($x$).  Use of these forms allows for a very simple $\chi^2$
type treatment of the band-power data---without the bias.

While we have found these approximations to the likelihood functions to
be quite adequate for dealing with the data we have explored to date,
and have given quite general arguments for why the tails behave as they
do, our checks have not been exhaustive. For example, some quoted CMB
anisotropy results are skewed to lower rather than higher amplitudes
(perhaps due to fitting out foregrounds or systematics), a situation
that the offset lognormal cannot fit. Just as one computes the curvature
about the maximum likelihood, so one can consider computing a skewness
that would encapsulate such behavior, but we will leave the search for
further likelihood function approximations to further exploration.

We have shown that the offset lognormal approximation applied to a
two-parameter ($\sigma_8$ and $n_s$) family of CDM models works very
well. We have also used this form to find the maximum-likelihood binned
power spectrum, given the band-power data.  The resulting graphs provide
a visual representation of the power spectrum constraints that is, in
our opinion, far superior to plotting all the band-power data on top of
each other.

The exercise of estimating the binned power spectrum immediately
raises the question of how well it would work to estimate
cosmological parameters using these as a ``super-radically compressed
dataset''.  We plan to pursue this question in future work.

Although our examples have focused on using our approximations in order
to derive parameter constraints from more than one dataset, we believe
they may also prove useful for estimating cosmological parameters
from single, very powerful datasets, such as those that are expected to
come from a number of experiments over the next decade.  We must note
though that once a dataset has sufficient ``spectral resolving power''
and dynamic range there is another approach that can be used to remove
the cosmic bias.  This alternative approach was suggested in
\citet{bjkpspec} and \citet{uroscopy}
and successfully applied to simulated MAP data in
\citet{OhSpergelHinshaw}.  The idea is to exploit the
fact that we expect there to be no fine features in the CMB power
spectrum and therefore use some smoothed version of the estimated power
spectrum to calculate the Fisher matrix.  Heuristically one expects this
to remove the bias, since upward-fluctuating points no longer receive
less weight than downward-fluctuating points.  Although this smoothing
technique is quite likely to be successful, we point out that, unlike
our ansatz, it relies on an assumption of the smoothness of the power
spectrum.

One of our main objectives with this paper is to provide observers
with a method for presenting their results that will allow efficient
combination with the results of others in order to create a joint
determination of cosmological parameters.  The method is fully
described in Appendix~\ref{app:recipe}.  In this appendix we also discuss
complications due to overlapping sky coverage and upper limits.

\acknowledgements AHJ would like to thank the members of the COMBAT
collaboration, especially P.G.\ Ferreira, S.\ Hanany and J.\ Borrill,
for discussions and advice. LK would like to thank Andrew Hamilton
for a useful conversation.  AHJ acknowledges support by NASA grants
NAG5-6552 and NAG5-3941 and by NSF cooperative agreement AST-9120005.

\appendix
\section{Signal-to-Noise Eigenmodes and the general form of the likelihood}
\label{app:snmode}

We start with the full-sky likelihood, either in the form of
Eq.~\ref{eqn:fullskylike} or in terms of $Z=\ln(\C+x)$,
Eq.~\ref{eqn:equalindepmode2}. We transform this form to signal-to-noise
eigenmodes ({\sl e.g.}, \citet{Bond94};\citet{BunnWhite};\citet{bjkpspec}).  Here, the data in
the signal-to-noise eigenmode basis are $d_k$, with diagonal covariance,
\begin{equation}
  \label{eq:snmodedef}
  \langle d_k d_{k'} \rangle = (1 + \sigma^2_{\rm th}\lambda_k)\delta_{kk'}.
\end{equation}
We allow the amplitude, $\sigma^2_{\rm th}$, for the signal contribution
to vary, since the eigenmodes only depends upon the shape of the signal
covariance, itself dependent on the input power spectrum.  If we define
the signal-to-noise transformation with power spectrum $\C_\ell^{\rm
  shape}$, then Eq.~\ref{eq:snmodedef} is valid for all $\C_\ell =
\sigma^2_{\rm th}\C_\ell^{\rm shape}$.  The eigenvalue for mode $k$ is
$\sigma_{\rm th}^2\lambda_k$, with units of (signal-to-noise)${}^2$.

Because the Gaussian variables, $d_k$, are statistically independent, the
likelihood is made up of independent contributions,
\begin{eqnarray}
  \ln {\cal L}/{\widehat{\cal L}} &=& -{{1\over2}}  \sum_k
  \bigg[{d_k^2 \over 1+{\widehat\sigma^2_{\rm th}} \lambda_k} {({\widehat\sigma^2_{\rm th}} -\sigma^2_{\rm th})\lambda_k \over 1+\sigma^2_{\rm th}
    \lambda_k} \nonumber \\
  && \quad + \ln{1+\sigma^2_{\rm th}
    \lambda_k \over 1+{\widehat\sigma^2_{\rm th}} \lambda_k}\bigg]\, , 
\label{eqn:StoNlike}
\end{eqnarray}
where a ``hat'' refers to the quantity at the likelihood maximum.

Introducing the number of modes with a given value of $\lambda$,
$g(\lambda )$, and the cumulative number of modes
$G(\lambda ) = \sum_{\lambda^\prime > \lambda} g(\lambda^\prime ) $,
this can be written in the suggestive form
\begin{eqnarray}
  \ln {\cal L}/{\widehat{\cal L}} &=& -{1\over2} \int dG(\lambda )\bigg[
  \left[Z(\lambda) - {\widehat Z}(\lambda)\right] \nonumber \\
  && + {\chi^2_\lambda\over 1 + {\widehat\sigma^2_{\rm th}} \lambda}
  \left(e^{-\left[Z(\lambda) - {\widehat Z}(\lambda)\right]}-1\right)\bigg]\, .
  \label{eqn:dGStoNlike}
\end{eqnarray}
Here $\chi^2_\lambda = \sum_{\lambda_k=\lambda} d_k^2 /g(\lambda)$ is
the $\chi^2$ per degree of freedom for modes of the $\lambda$. On
average, $\chi^2_\lambda$ approaches $(1 + {\widehat\sigma^2_{\rm th}}
\lambda)$, in which case the form is a sum of terms like
Eq.~\ref{eqn:equalindepmode}. The variables $Z(\lambda) \equiv \ln
(1+\sigma^2_{\rm th} \lambda )$ are analogous to the form we have been
using if $\lambda^{-1}$ is interpreted as a special case of the offset
$x$.  The integral is to be interpreted in the Stieltjes sense, as a sum
over the discrete $\lambda$ spectrum, $ \sum_\lambda g(\lambda) (\ldots)$.

Consider what happens asymptotically with increasing $\sigma^2_{\rm th}$.
The modes with $\sigma^2_{\rm th} \lambda > 1$ contribute, so
$G(\sigma_{\rm th}^{-2})\ln \sigma^2_{\rm th}$ is the leading behavior. As
$\sigma^2_{\rm th}$ goes up, more eigenmodes may contribute, $ \ln {\cal
  L}/{\widehat{\cal L}} $ decreases faster, modifying the tail.

\section{Data Reporting Recipe}
\label{app:recipe}

We recommend reporting future (and, if possible, past) CMB results in a
form that will render them amenable to this ``radically-compressed''
analysis. Thus, experimenters and phenomenologists ought to provide
estimates of
\begin{itemize}
\item $\C_\ell\equiv\ell(\ell+1)C_\ell/(2\pi)$, the power spectrum in
  appropriate bins;
\item ${\cal F}^{-1}_{\ell\ell'}$, the curvature or covariance matrix of the
  power spectrum estimates; and
\item $x_\ell$, the quantity such that $Z_\ell\equiv\ln({\cal C}_\ell+x_\ell)$
  is approximately distributed as a Gaussian.
\end{itemize}

We here provide an outline of the steps needed to provide the
appropriate information.  Current listings of publicly-available results
will be posted at {\tt{http://www.cita.utoronto.ca/$\sim$knox/radical.html}}.
Please contact the authors to have results included.
\\

\noindent{1)~\it Divide the power spectrum into discrete bins.}\/  To prevent
significant loss of shape information, the bins should not be too large.
However, there may be a problem with making the bins too small.  The closer
we are to the case of well-determined, independent bins, the better our
ansatz is expected to work.  Thus bins should be large enough to keep
relative error bars smaller than 100\% and bin to bin correlations small.
The shape-dependence resulting from coarse bins can be reduced by the use of 
a filter function for each band (see \citet{Knox99}), as was done with
the MSAM (Medium Scale Anisotropy Measurement) three-year data 
\citep{wilson99}.
\\
\noindent{2)~\it Find the power in each bin that maximizes the likelihood
  and evaluate the curvature matrix at this point.}\/ This can be done
using your favorite likelihood search algorithm.  For COBE/DMR and
Saskatoon we have used the iterative scheme described in
\citet{bjkpspec}.  Our current implementation does not include a
transformation to S/N-eigenmode space. In a forthcoming paper
\citep{knoxjaffequadest}, we will provide detailed information on the
implementation of this quadratic estimator and an appropriate sample set
of programs.
\\
\noindent{3)~\it Estimate $x$ for each of the bins.}\/  If the
likelihood is calculated explicitly, this can simply be done by
numerical fitting to the functional form of our ansatz,
Eqs.~\ref{eqn:gausslike} and following, or Eq.~\ref{eqn:bandxb}.  If the
likelihood peak is determined by the iterative quadratic scheme or some
other method which also calculates the curvature matrix (or,
less-preferably, Fisher matrix), the appropriate formulae from
Sec.~\ref{sec:solution} (for total-power mapping experiments) or
Sec.~\ref{sec:chop} (Eq.~\ref{eqn:getxb2} for chopping experiments).
\\
\noindent{4)}~Do not alter the curvature matrix by folding in the
calibration uncertainty in any way.  Report the calibration uncertainty
separately.

\subsection{Special Cases}

\noindent{1)~\it Overlapping sky coverage.}\/ Power spectrum
constraints will be correlated if they are from datasets with
overlapping sky coverage and sensitivity to similar angular scales.
We have no general theory of these correlations.  Proper combination
of overlapping datasets appears to require a joint likelihood analysis
to produce their combined constraints on the power spectrum.

\noindent{2)~\it Upper Limits.}\/
For datasets that can only provide an upper limit to the power spectrum
amplitude, a simple option would be to calculate the full likelihood
directly and simply fit to one of the two forms for $\C_\ell\ge0$ but with a
negative or very small $\C_\ell$ (and such that $\C_\ell+x_\ell>0$).
This is what we have done for Fig.~\ref{fig:ovrolike}
which demonstrates that both of our approximate forms
work fairly well, especially the full form of Section~\ref{sec:indmode}.

Although the results in Fig.~\ref{fig:ovrolike} look quite impressive,
they say nothing about how well the window function
tells us which regions of the power spectrum are being constrained
by the data.  In other words, does the trace of the window function
make a good filter function?  Therefore, we present the
following alternative method for reporting upper limits which
includes a prescription for creation of a filter function.

The data can be reported as amplitudes of signal-to-noise eigenmodes and
their eigenvalues (see Appendix~\ref{app:snmode}).  One need only report
the modes with the largest eigenvalues.  The number of modes that it is
necessary to report is likely to be quite small.  The likelihood of
$\sigma^2_{\rm th}$, where $\C_\ell= \sigma^2_{\rm th}\C_\ell^{\rm
  shape}$, is then:
\begin{equation}
  \chi^2 \equiv -2\ln{\cal L} = \sum_i\left[\ln\left(1+\sigma^2_{\rm th}
      \lambda_i\right) +
    {D_i^2\over1+\sigma^2_{\rm th} \lambda_i}\right]
\end{equation}
where $D_i$ is the amplitude of the $i$th mode, $\lambda_i$ is its eigenvalue,
and $\C_\ell^{\rm shape}$ is the power spectrum used to define the S/N-modes.
Of course, we want the likelihood to be a function of, {\sl e.g.}, a binned power
spectrum.  It is, via:
\begin{equation}
  \sigma^2_{\rm th}={1\over \C} \sum_B f_B \C_B
\end{equation}
where we have assumed a flat power spectrum ($\C_\ell^{\rm
  shape}=\C={\rm const}$),
$f_B$ is related to the window function as in Eq.~\ref{eqn:win2filt} or,
better, derived from the Fisher matrix as described in \citet{bjkpspec}.
It is straightforward to calculate the derivatives of this $\chi^2$ with
respect to $\C_B$ in order to combine upper limits with detections
and perform the search procedure described in Section~\ref{sec:bands}.

\section{Band Powers}
\label{app:Bandpowers}

\begin{deluxetable}{lrrrrl}
  \scriptsize
  \rotate\tablecolumns{6}
  \tablecaption{Input band powers, standard errors, and noise
    contributions to the variance ($x$) in $\muK^2$. \label{tab:Bandpowers}}
  \tablehead{
    \colhead{dataset} & \colhead{$\ell_{\rm -}$--$\ell_{\rm +}$} & \colhead{BP} & 
    \colhead{err} & \colhead{$x$} & reference
    }
  \startdata
firs           &    4--25&  927.8&  440.7&   ?? & \citet{firs}; \citet{Bond94}\\
tenerife595    &   13--28& 1164  &  705.9&   ?? & \tablenotemark{a}\\
BAM            &   31--90&  870.3&  478.5&   ?? & \citet{BAM}\\
SP91-6225-63   &   35--98&  892.2&  382.8&   ?? & \citet{sp91}; \citet{sp91b}\\
SP94-62-4ch    &   33--95&  837.5&  384.6&   ?? & \citet{sp94}\\
SP94-62-3ch    &   40--114& 1632  &  584.5&   ?? & \citet{sp94}\\
pyth96I-II-III &   52--99& 2916  & 1351  &   ?? & \citet{python123}; \citet{python2}\\
pyth96III      &  132--237& 3364  & 1565  &   ?? & \citet{python123}; \citet{python2}\\
qmap\_q        &   79--143& 2704  &  520.0&   ?? & \tablenotemark{b}\\
qmap\_ka1      &   60--101& 2209  &  604.5&   ?? & \tablenotemark{b}\\
qmap\_ka2      &   99--153& 3481  &  760.5&   ?? & \tablenotemark{b}\\
toco97\_3      &   45--81&  1600  &  720  &  500 & \citet{mat97}\\
toco97\_4      &   70--108& 2040  &  600  &  100 & \citet{mat97}\\
toco97\_5      &   90--138& 4900  &  850  &    0 & \citet{mat97}\\
toco97\_6      &  135--180& 7850  & 1300  &    0 & \citet{mat97}\\
toco97\_7      &  170--237& 7170  & 1300  & 3000 & \citet{mat97}\\
sp89           &   87--247&    0.0& 1459  & 1830 & \citet{sp89}\\
argo           &   69--144& 1060  &  613.0&   ?? & \citet{argo}; \citet{argo2}\\
MAX4av         &   89--249& 2586  &  876.9&   ?? & \citet{MAX4}; \citet{MAX4b}\\
MAX5av         &   89--249& 1511  &  573.8&   ?? & \citet{MAX5}; \citet{MAX5}\\
ovro22         &  362--759& 3127  &  813.1&   ?? & \citet{ELthesis}\\
cat1           &  349--473& 2583  & 1512  &   ?? & \citet{catold}\\
cat2           &  559--709& 2401  & 1584  &   ?? & \citet{catold}\\
cat1-98        &  349--473& 3937  & 2322  &15700 & \citet{cat98}\\ 
cat2-98        &  559--709&    0.0& 5031  &15700 & \citet{cat98}\\
OVRO           & 1147--2425&   72.4&  380.3&  367 & \citet{OVRO}\\
SuZIE          & 1366--3000&  354.3&  753.4&  122\tablenotemark{s} & \citet{SuZIE}
\enddata
\tablenotetext{a}{\citet{tenerife}; \citet{tenerife2};
  \citet{tenerife3}}
\tablenotetext{b}{\citet{qmap1}; \citet{qmap2}; \citet{qmap3}}
\tablenotetext{s}{$x=419$ is a better fit for the equal variance form
  (not used in the calculations of Section~\ref{sec:bands}).}
\tablecomments{These data, their
    window functions, and also data for MSAM-3yr \citep{wilson99},
    Saskatoon \citep{nett95} and COBE/DMR \citep{DMR} are available at
    http://www.cita.utoronto.ca/$\sim$knox/radical.html. These last three
    datasets are not shown in the table here because they have multiple
    bands with covariance matrices.}
\end{deluxetable}

The numbers in Table~\ref{tab:Bandpowers} were used to form part of the
weight matrix in
Eq.~\ref{eqn:chisq}:  $W_{ij}=1/\sigma_i^2 \delta_{ij}$ where $\sigma_i$
is from the ``standard error'' column of the Table.  These
standard errors are derived from published likelihood maxima ($D_i$),
68\% confidence upper limits ($D_i^u$) and lower limits ($D_i^l$).
Since the upward and downward excursions from the mode to the upper and
lower limits are usually different, there is some freedom in assigning a
single standard error.  We define $\sigma_i$ as an average of these
excursions:
\begin{equation}
\sigma_i =  \left[ \left(D_i^u-D_i\right) +\left(D_i-D_i^l\right)
\right]/2.
\end{equation}
If the published number is linear instead of quadratic, then $D_i =
\delta T_i^2$, etc.\  and the above equation still applies.  We have
also tried producing $\sigma_i$ from averaging the inverse square of
the upward and downward deviations, and found no significant
difference in the results (power in bands changes by less than 10\% of
the error bar).

We also found not much difference in the results depending on how we
treated calibration uncertainty.  Most experiments report their upper
and lower limits with calibration uncertainty included.  Only for
Saskatoon, MSAM, QMAP, TOCO97 and TOCO98 have we included calibration
uncertainty by treating it as an independent parameter ($u_\alpha$ in
Eq.~\ref{eqn:chisq}).

Missing from the table are detections from the
White Dish \citep{whitedish} experiment.  The White Dish dataset was
compressed
to two band-power detections with sensitivity in the range
$\ell\sim300$ to $\sim 600$.  A recent reanalysis \citep{wdnewanalysis},
results in upper limits which are sufficiently loose that including
them would make no difference in our power spectrum determination.
Both these analyses use only a small subset of the available data; a
complete analysis will probably provide detections \citep{wdinprep}.


\begin{thebibliography}{}
\bibitem[Bond, Efstathiou \& Tegmark(1997)]{BET97} Bond, J.R.,
  Efstathiou, G., and Tegmark, M.\ 1997, \mnras, {291}, L33.
\bibitem[Allen \etal(1997)]{allenetal} Allen, B.\ \etal1997, \prl,
  {79}, 2624
\bibitem[Baker \etal(1998)]{cat98} Baker, J.C.\ \etal1998, \mnras,
  submitted.
\bibitem[Bartlett \etal(1999)]{bartlettetal} Bartlett, J.G., Douspis,
  M., Blanchard, A., Le Dour, M., 1999, \aap,
  submitted. astro-ph/9903045.
\bibitem[Bennett \etal(1996)]{DMR} Bennett, C.L., Banday, A.J., K.M.\
  Gorski, Hinshaw, G., Jackson, P.D., Keegstra, P., Kogut, A., Smoot,
  G.F., Wilkinson, D., and Wright, E.L.\ 1996, \apjl, {464}, L1, and
  4-year COBE/DMR references therein.
\bibitem[Bond(1994)]{Bond94} Bond, J.R.\ 1994, \prl, {74}, 4369.
\bibitem[Bond \& Jaffe(1998a)]{bondjaffe1} Bond, J.R.\ and Jaffe, A.H.\
  1998a, in {Microwave Background Anisotropies, Proceedings of the XVI
    Rencontre de Moriond}, ed.\ Bouchet, F.R. (Paris: Editions Frontieres).
\bibitem[Bond \& Jaffe(1998b)]{bondjaffe2} Bond, J.R.\ and Jaffe, A.H.\
  1998b, Phil.\ Trans.\ R.\ Soc.\ Lond.\ A, to appear.
\bibitem[Bond, Jaffe \& Knox(1998)]{bjkpspec} Bond, J.R., Jaffe, A.H.\
  and Knox, L.E.\ 1998, \prd, in press; astro-ph/9708203.
\bibitem[Bond, Pogosyan \& Souradeep(1998)]{BPS98} Bond, J.R., Pogosyan,
  D., and Souradeep, T.\ 1998, {Class.\ Quant.\ Gravity}, in press.
\bibitem[Bunn \& Sugiyama(1995)]{BunnSugiyama} Bunn, E.F.\ and Sugiyama,
  N.\ 1995, \apj, {446}, 49.
\bibitem[Bunn \& White(1997)]{BunnWhite} Bunn, E.F.\ and White, M.\ 1997,
  \apj, {480}, 6.
\bibitem[Cheng \etal(1994)]{msam92} Cheng, E.S.\ \etal1994, \apjl,
  {422}, L37.
\bibitem[Cheng \etal(1997)]{msam95} Cheng, E.S.\ \etal1997, \apjl,
  {488}, L59.
\bibitem[Church \etal(1997)]{SuZIE} Church, S.E.\ \etal1997, \apj,
  {4 84}, 523.
\bibitem[Clapp \etal(1994)]{MAX4} Clapp, A.C.\ \etal1994, \apjl,
  {433}, L57. 
\bibitem[DeBernardis \etal(1994)]{argo2} DeBernardis, P.\ \etal1994,
  \apjl, {422}, L33.
\bibitem[de Oliveira-Costa \etal(1998)]{qmap3} de Oliveira-Costa, A.\
  \etal1998, astro-ph/9808045.
\bibitem[Devlin \etal(1994)]{MAX4b} Devlin, M.J.\ \etal1994, \apjl,
  {430}, L1.
\bibitem[Devlin \etal(1998)]{qmap1} Devlin, M.\ \etal1998,
  astro-ph/9808043.
\bibitem[Efstathiou \etal(1998)]{efstetal98} Efstathiou, G., Bridle,
  S.L., Lasenby,  A.N., Hobson, M.P.\ and Ellis, R.S.\ 1997, \mnras,
  submitted. 
\bibitem[Ferreira, Magueijo \& Gorski(1998)]{weird16} Ferreira, P.G.,
  Magueijo, J., and Gorski, K.M.\ 1998, \apjl, accepted; astro-ph/9803256.
\bibitem[Gaier \etal(1991)]{sp91} Gaier, T.\ \etal1992, \apjl, {398},
  L1.
\bibitem[Ganga \etal(1994)]{firs} Ganga, K., Page, L., Cheng, E.S.,
  Meyer, S.S.\ 1994, \apjl, {432}, L15.
\bibitem[Griffin \etal(1998)]{wdinprep} Griffin, G.S., Nguy\^en, H.T.,
  Peterson, J.B., and Tucker, G.S.\ 1998, in preparation.
\bibitem[Guitteriez de la Cruz \etal(1995)]{tenerife} Gutteriez de la
  Cruz, S.M.\ \etal, \apj, {442}, 10.
\bibitem[Gundersen \etal(1993)]{MAX3} Gundersen, J.O.\ \etal1993, \apjl,
{413}, L1.
\bibitem[Gundersen \etal(1995)]{sp94} Gundersen, J.O.\ \etal1995,
\apjl, {443}, L57.
\bibitem[Hamilton(1997a)]{hamilton} Hamilton, A.J.S.\ 1997a,
  astro-ph/9701008.
\bibitem[Hamilton(1997b)]{hamilton97b} Hamilton, A.J.S.\ 1997b,
  astro-ph/9701009.
\bibitem[Hancock \etal(1998)]{hancockrocha} Hancock, S., Rocha, G.,
  Lasenby, A.N., and Gutierrez, C.M.\ 1998, \mnras, {294}, L1.
\bibitem[Hancock \etal(1994)]{tenerife2} Hancock, S.\ \etal1994,
  Nature, {367}, 333.
\bibitem[Herbig \etal(1998)]{qmap2} Herbig, T.\ \etal1998,
  astro-ph/9808044.
\bibitem[Hobson \& Magueijo(1996)]{HobsonMagueijo} Hobson, M.P.\ and
  Magueijo, J.\ 1996, \mnras, {283}, 1133.
\bibitem[Jaffe, Knox \& Bond(1998)]{jkbtexas} Jaffe, A.H., Knox, L., and
  Bond, J.R.\ 1998, in {Proceedings of the Eighteenth Texas Symposium on
    Relativistic Astrophysics}, ed.\ Frieman, J., Olinto, A., and Schramm,
  D., in press.
\bibitem[Jungman \etal(1996)]{Jungmanetal} Jungman, G., Kamionkowski,
  M., Kosowsky, A., and Spergel, D.N.\ 1996, \prl, {76}, 1007.
\bibitem[Kneissel \& Smoot(1993)]{GneSmoot} Kneissl, R.\ and Smoot, G.\
  1993, COBE note 5053.
\bibitem[Knox(1999)]{Knox99} Knox, L.\ 1999, \prd, in press.
\bibitem[Knox(1995)]{Knox95} Knox, L.\ 1995, \prd, {52}, 4307.
\bibitem[Knox \& Jaffe(1998)]{knoxjaffequadest} Knox,
  L.E.\ and Jaffe, A.H.\ 1998, in preparation.
\bibitem[Leitch(1998)]{ELthesis} Leitch, E.\ 1998, Caltech PhD.\ thesis.
\bibitem[Lim \etal(1996)]{MAX5} Lim, M.A.\ \etal1996, \apjl, {469}, L69.
\bibitem[Lineweaver(1997)]{lineweaver97} Lineweaver, C.H.\ 1997,
  astro-ph/9702042.
\bibitem[Lineweaver(1998a)]{lineweaver98a} Lineweaver, C.H.\ 1998a, \apjl,
submitted; astro-ph/9805326.
\bibitem[Lineweaver(1998b)]{lineweaver98b} Lineweaver, C.H.\ 1998b,
  astro-ph/9803100.
\bibitem[Lineweaver(1998c)]{lineweaver98c} Lineweaver, C.H.\ 1998c,
  astro-ph/9801029.
\bibitem[Lineweaver \& Barbosa(1998)]{lineweaverBarbosa97} Lineweaver,
  C.H.\ and Barbosa, D.\ 1998, \apj, {496}.
\bibitem[Lineweaver \& Smoot(1993)]{LineSmoot} Lineweaver, C.\ and Smoot,
  G.\ 1993, COBE note 5051.
\bibitem[Masi \etal(1996)]{argo} Masi, S.\ \etal1996, \apjl, {463}, L47.
\bibitem[Meinhold \& Lubin(1989)]{sp89} Meinhold, P.\ and Lubin, P.\
  1989, \apjl, {370}, L11.
\bibitem[Miller \etal(1999)]{mat98}Miller, A.D.\etal1999, \apjl,
  astro-ph/9906421. 
\bibitem[Myers, Readhead \& Lawrence(1993)]{OVRO} Myers, S.T.,
  Readhead, A.C.S., and Lawrence, C.R.\ 1993, \apj, {405}, 8.
\bibitem[Netterfield \etal(1997)]{nett95} Netterfield, C.B.,
  Devlin, M.J., Jarosik, N., Page, L., and Wollack, E.J.\ 1997, \apj,
  {474}, 47.
\bibitem[Oh, Spergel \& Hinshaw(1998)]{OhSpergelHinshaw} Oh, S.P.,
  Spergel, D.N., and Hinshaw, G.\ 1998, astro-ph/9805339.
\bibitem[Platt \etal(1997)]{python123} Platt, S.R., Kovac, J., Dragovan,
  M., Peterson, J.B., and Ruhl, J.E.\ 1997, \apjl, {475}, L1.
\bibitem[Ratra \etal(1997)]{wdnewanalysis}Ratra, B.\ \etal1997,
  astro-ph/9710270.
\bibitem[Ruhl \etal(1995)]{python2} Ruhl, J.E.\ \etal1995, \apjl,
  {453}, L1.
\bibitem[Schuster \etal(1991)]{sp91b}Schuster, J.\ \etal1993, \apjl,
  {412}, L47.
\bibitem[Scott \etal(1996)]{catold} Scott, P.F.S.\ \etal1996, \apjl,
  {461}, L1.
\bibitem[Seljak(1997)]{uroscopy} Seljak, U.\ 1997, astro-ph/9710269.
\bibitem[Tanaka \etal(1996)]{MAX5b} Tanaka, S.T.\ \etal1996, \apjl,
  {468}, L81.
\bibitem[Tegmark(1997)]{tegmark} Tegmark, M.\ 1997, \prd, {55},
5895; astro-ph/9611174.
\bibitem[Tegmark(1998)]{tegmark98} Tegmark, M.\ 1999, \apjl, {514}, L69.
\bibitem[Tegmark \& Hamilton(1998)]{TegHam} Tegmark, M.\ and Hamilton,
  A.\ 1998, in {Proceedings of the Eighteenth Texas Symposium on
    Relativistic Astrophysics}, ed.\ Frieman, J., Olinto, A., and Schramm,
  D., in press.
\bibitem[Tegmark, Taylor \& Heavens(1997)]{TTHklmode} Tegmark, M.,
  Taylor, A., and Heavens, A.\ 1997, \apj, {480}, 22.
\bibitem[Torbet \etal(1999)]{mat97} Torbet, E.\ \etal1999, astro-ph/9905100.
\bibitem[Tucker \etal(1993)]{whitedish} Tucker, G.S., Griffin, G.S.,
  Nguy\^en, H.T., and Peterson, J.B.\ 1993, \apjl, {419}, L45.
\bibitem[Tucker \etal(1997)]{BAM} Tucker, G.S., Gush, H.P., Halpern, M.,
  Shinkoda, I., and Towlson, W.\ 1997, \apjl, {475}, L73.
\bibitem[Watson \etal(1992)]{tenerife3} Watson, R.\ \etal1992, Nature,
  {357} 660.
\bibitem[Wilson \etal(1999)]{wilson99} Wilson, G.\ \etal1999, astro-ph/9902047.
\bibitem[Wollack \etal(1997)]{woll95} Wollack, E.J., Devlin, M.J.,
  Jarosik, N., Netterfield, C.B., Page, L., and Wilkinson, D.\ 1997, \apj,
  {476}, 440.
\end{thebibliography}
\end{document}